\documentclass[fleqn,usenatbib]{mnras}

\usepackage{newtxtext,newtxmath}

\usepackage[T1]{fontenc}
\usepackage{ae,aecompl}
\usepackage{xspace}

\usepackage{graphicx}	
\usepackage{amsmath}	
\usepackage{amssymb}	
\usepackage[percent]{overpic}

\newcommand{\msun}{\,M$_{\odot}$}

\newcommand{\kms}{\,km\,s$^{-1}$\xspace}
\newcommand{\kmspc}{\,km\,s$^{-1}$\,pc$^{-1}$\xspace}

\newcommand{\cmc}{\,cm$^{-3}$\xspace}

\renewcommand{\deg}{\ensuremath{^{\circ}}\xspace}
\newcommand{\pc}{\,{\rm pc}}
\newcommand{\kpc}{\,{\rm kpc}}

\newcommand{\arepo}{\textsc{Arepo}\xspace}

\newcommand{\disperse}{DisPerSE\xspace}

\title[The Cloud Factory I: Generating resolved filamentary molecular clouds from galactic-scale forces]{The Cloud Factory I: Generating resolved filamentary molecular clouds from galactic-scale forces}

\author[Smith et al.]{
Rowan J. Smith,$^{1}$\thanks{E-mail: rowan.smith@manchester.ac.uk}
Robin G. Tre{\ss},$^{2}$
Mattia C. Sormani,$^{2}$
Simon C.O. Glover,$^{2}$
\newauthor
Ralf S. Klessen,$^{2,3}$
Paul C. Clark,$^{4}$
Andr\'es F. Izquierdo,$^{1}$
Ana Duarte Cabral,$^{4}$
\newauthor
Catherine Zucker$^{5}$
\\
$^{1}$ Jodrell Bank Centre for Astrophysics, Department of Physics and Astronomy, University of Manchester, Oxford Road, Manchester M13 9PL, UK\\
$^{2}$Universit\"{a}t Heidelberg, Zentrum f\"{u}r Astronomie, Institut f\"{u}r theoretische Astrophysik, Albert-Ueberle-Str. 2, 69120 Heidelberg, Germany \\
$^{3}$Universit{\"a}t Heidelberg, Interdisziplin{\"a}res Zentrum f{\"u}r Wissenschaftliches Rechnen, INF 205, 69120 Heidelberg, Germany\\
$^{4}$ School of Physics and Astronomy, Queens Buildings, The Parade, Cardiff University, Cardiff, CF24 3AA\\
$^{5}$ Harvard Astronomy, Harvard-Smithsonian Center for Astrophysics, 60 Garden St., Cambridge, MA 02138, USA
}

\date{Accepted XXX. Received YYY; in original form ZZZ}

\pubyear{2015}

\begin{document}
\label{firstpage}
\pagerange{\pageref{firstpage}--\pageref{lastpage}}
\maketitle

\begin{abstract}
We introduce  a new suite of simulations, "The Cloud Factory", which self-consistently forms molecular cloud complexes at high enough resolution to resolve internal substructure (up to 0.25 \msun \ in mass) all while including galactic-scale forces. We use a version of the \arepo code modified to include a detailed treatment of the physics of the cold molecular ISM, and an analytical galactic gravitational potential for computational efficiency. The simulations have nested levels of resolution, with the lowest layer tied to tracer particles injected into individual cloud complexes. These tracer refinement regions are embedded in the larger simulation so continue to experience forces from outside the cloud. This allows the simulations to act as a laboratory for testing the effect of galactic environment on star formation. Here we introduce our method and investigate the effect of galactic environment on filamentary clouds. We find that cloud complexes formed after a clustered burst of feedback, have shorter lengths and are less likely to fragment compared to quiescent clouds (e.g.\ the Musca filament) or those dominated by the galactic potential (e.g.\ Nessie). Spiral arms and differential rotation preferentially align filaments, but strong feedback randomises them. Long filaments formed within the cloud complexes are necessarily coherent with low internal velocity gradients, {which has implications for the formation of filamentary star-clusters}. Cloud complexes formed in regions dominated by supernova feedback have fewer star-forming cores, and these are more widely distributed. These differences show galactic-scale forces can have a significant impact on star formation within molecular clouds.
\end{abstract}




\section{Introduction}


Galactic dynamics and star formation are inextricably linked. 
Galactic-scale structures, such as spiral arms or bars, aggregate cold molecular gas, differential rotation stretches it, and feedback from supernovae injects momentum into the interstellar medium (ISM), driving the gas apart again. All of these factors have a profound effect on the mass, thermodynamical state and velocity structure of the resulting molecular clouds, and hence on the ability of the gas to fragment into stars and stellar clusters. However, a detailed understanding of how galactic-scale dynamics influences molecular cloud substructure and fragmentation remains elusive. In this paper we seek to link these scales by introducing the `Cloud Factory', a new suite of simulations with sufficient dynamical range to model the behaviour of the ISM in a typical spiral galaxy on scales ranging from the entire galaxy down to individual filaments and clumps within selected molecular clouds.
This acts as a unique laboratory to test how the large-scale galactic environment influences the local star formation process.

Of particular interest in the literature has been the role of turbulence in cloud fragmentation. Several studies have used results from idealized simulations of interstellar turbulence -- specifically, the finding that the gas develops a log-normal density probability distribution function (PDF) with a width related to the properties of the turbulence -- in combination with a model for the onset of gravitational collapse to predict the efficiency with which gas is transformed into stars \citep[e.g][]{Padoan02, Krumholz05, Hennebelle08,Federrath13,Burkhart18}. Simulations of this process typically use a periodic box setup where turbulence is driven at large scales to generate turbulent scaling laws reminiscent of those observed in molecular clouds \citep{Larson81,Heyer04}. Other authors have used simulations of decaying turbulence \citep[e.g.][]{MacLow98, Klessen01,Bonnell06,Vazquez-Semadeni07,Smith09b} to look at the assembly of massive stars in bound collapsing clusters. However, there remains the question of how closely real molecular clouds resemble these idealised models.

Furthermore, it is not just the global turbulent velocity field that seems to play a role in star formation within molecular clouds, but also the morphology of the gas. Nearby clouds observed in dust emission break down into networks of filamentary structures  \citep[e.g.][]{Andre10,Menshchikov10,Arzoumanian11,Schneider12}. Similar structures are also observed in C$^{18}$O and $^{13}$CO emission \citep{Panopoulou14, Suri19}, which appear to decompose into smaller scale `fibers' when a higher critical density tracer is used \citep{Hacar13,Hacar16,Henshaw16}

 Filament fragmentation differs from 3D Jeans fragmentation \citep{Jeans1902} (usually considered in spherical symmetry for simplicity) in that it occurs above a critical line mass \citep{Larson73c,Larson85,Inutsuka92,Inutsuka97}, takes place over a longer time-scale \citep{Pon12}, and occurs on a characteristic length scale \citep{Larson85}. In addition to determining fragmentation within the cloud, filaments may enhance the accretion onto cores at the `hubs' where they intersect, enabling the assembly of the high mass end of the stellar initial mass function \citep{Smith11a,Myers11,Peretto12,Smith16}. These filament networks must originate during the formation of the clouds since they are seen from the lowest densities \citep{Arzoumanian11}. It is therefore  crucial to move our simulation efforts forward to a more self-consistent picture, where turbulence and filaments are generated self-consistently during cloud formation in a realistic galactic environment, rather than arbitrarily imposed in the initial conditions.

 Recent observations have shown that many clouds take the form of extremely long (100~pc or more) filaments, some of which seem to correlate with the dense centres of spiral arms (e.g.\ the `Nessie' filament; see \citealt{Goodman14} and \citealt{Zucker15}) while others may be more associated with inter-arm regions (e.g.\ Giant Molecular Filaments or GMFs; see \citealt{Ragan14} and \citealt{Abreu-Vicente16}). The properties of these filamentary clouds seem to vary as a function of galactic environment \citep{Zucker17}. 
 
Both arm and inter-arm filaments are characterised by a high degree of velocity coherence. For example, in the sample of \citet{Ragan14} the GMFs spanned velocity ranges between 5-13 \kms over lengths of 51-234 pc. On smaller scales within Giant Molecular Cloud complexes, filamentary clouds are also observed to be velocity coherent and have a low velocity dispersion when not dominated by feedback processes. The pristine `Musca' filament has subsonic velocity dispersions along its 6.5 pc length when observed in C$^{18}$O, and a velocity gradient of only 0.3 \kmspc  \citep{Hacar16} similar to the GMFs.

In order to better represent such clouds, much recent theoretical effort has gone into improving our numerical models of cloud formation. Of most relevance for this work are simulations which seek to study cloud formation within a galactic context. For example, \citet{Dobbs06a} simulated gas discs responding to a galactic spiral potential and showed that the majority of clouds were unbound \citep{Dobbs08b,Dobbs11}. \citet{Tasker09} studied gas clouds within a galaxy disc without spiral arms and showed that the cloud-cloud collision timescale was only a fifth of the orbital time and consequently that this was an efficient method of injecting turbulence into the gas. 

One can then zoom-in to galactic models to study the molecular gas at higher resolution. \citet{Smith14a} increased the gas mass resolution in $1/8$th of a spiral galaxy disc to only 4 \msun \ (cell radii of $\sim 0.3$ pc), to show that $\sim 40\%$ of the gas was CO dark, and that this dark gas extended outwards in long (100s of pc) filaments from clouds in inter-arm regions. \citet{Duarte-Cabral16} performed synthetic observations of clouds extracted from a zoomed in region (with SPH particle mass of $3.75 \, {\rm M_{\odot}}$, and typical SPH kernel mass of $125  \, {\rm M_{\odot}}$) from \citet{Dobbs15}, finding that filamentary clouds were predominantly formed in the inter-arm regions. \citet{Duarte-Cabral17} then followed up a selection of these giant filaments at 1 pc resolution to show that they formed predominantly through galactic shear, and are most defined at the bottom of the spiral potential well, but typically do not survive the crossing of the spiral arm as single filaments but merge into GMCs. In other high resolution zoom simulations, \citet{Butler15} extracted a kpc box from a galaxy model where clouds formed predominantly through cloud-cloud collisions and resolved the gas down to 0.1 pc scales, finding large velocity dispersions in contrast to observations.

Another approach to generating more realistic molecular clouds is to focus on large boxes containing turbulence driven by supernovae, either alone \citep[e.g.][]{Walch15,Ibanez-Mej16,Padoan16} or in combination with other forms of feedback \citep[e.g.][]{Gatto17,Peters17}. Simulations of stratified boxes have been used to investigate how feedback-driven turbulence drives the matter cycle of the ISM, and have shown that a combination of random and clustered supernova driving is needed to reproduce the properties of the ISM \citep{Gatto15,Girichidis16,Iffrig17,Hennebelle18}. Zoom-in simulations by \citet{Seifried17,Seifried18} have been used to study at high resolution the behaviour of several molecular clouds selected from the SILCC simulations of \citet{Walch15}, and have shown that   
supernova explosions are inefficient at driving turbulence within pre-existing dense molecular clouds. The \textsc{Tigress} simulations of \citet{KimCG13} and \citet{KimCG17} also included differential rotation into such boxes, with the resultant shear making the molecular clouds easier to destroy. 

The simulations mentioned above still suffer from some limitations. For example, the stratified boxes in all but one case neglect galactic shear, and none contain spiral arms. Galactic-scale simulations generally fail to resolve gas on sub-parsec scales without isolating the boxes from large scale-forces or adopting cooling prescriptions that forgo explicit modelling of cold molecular gas and the chemical phases of the ISM \citep{Renaud13,Bournaud15}. 

Resolving cold molecular gas at sub-pc scales is crucial for determining where dense star-forming clumps and CO molecules will form within the clouds \citep{Joshi19}. The `Cloud Factory' simulations we present here include: a galactic potential with differential rotation in the disc, spiral arms, clustered and random supernovae feedback, gas self-gravity, sink particles to represent star formation, and a time dependent chemical model, all while resolving the dense gas to sub-parsec scales. In our highest zooms we reach target mass resolutions of 0.25 \msun \ in selected clouds within the galaxy scale model, meaning for the first time the detailed dynamics and fragmentation within a molecular cloud can be linked to its galactic environment. Future work will include magnetic fields and a more sophisticated treatment of stellar feedback. In this first paper we will explore how galactic forces and supernova feedback shape the morphology and dynamics of filamentary molecular cloud structure.

This paper can be broadly divided into two main parts. In Section \ref{sec:methods} we outline the methodology of our Cloud Factory simulations that we will use for this and future works. In Section \ref{sec:results} we give an illustrative example of its power by investigating the properties of filamentary clouds formed via this method with and without clustered supernova feedback. After this we discuss our results in Section \ref{sec:discussion} and summarise our conclusions in Section \ref{sec:conclusions}.

\section{The Cloud Factory Simulations}
\label{sec:methods}

\begin{figure*}
\begin{center}
\begin{tabular}{c c}

\begin{overpic}[scale=0.32]{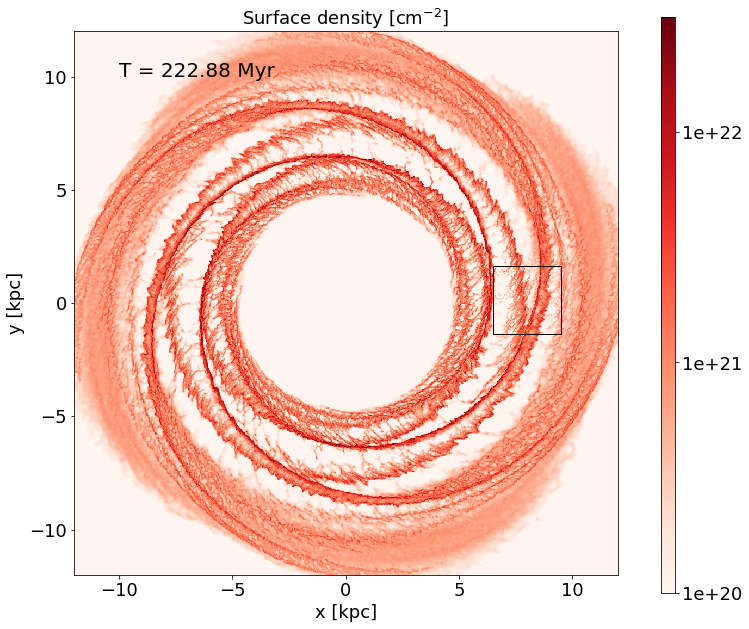}
\put (30,10) {\makebox(0,0){{\bf Potential dominated}}}
\end{overpic}
\includegraphics[width=\columnwidth]{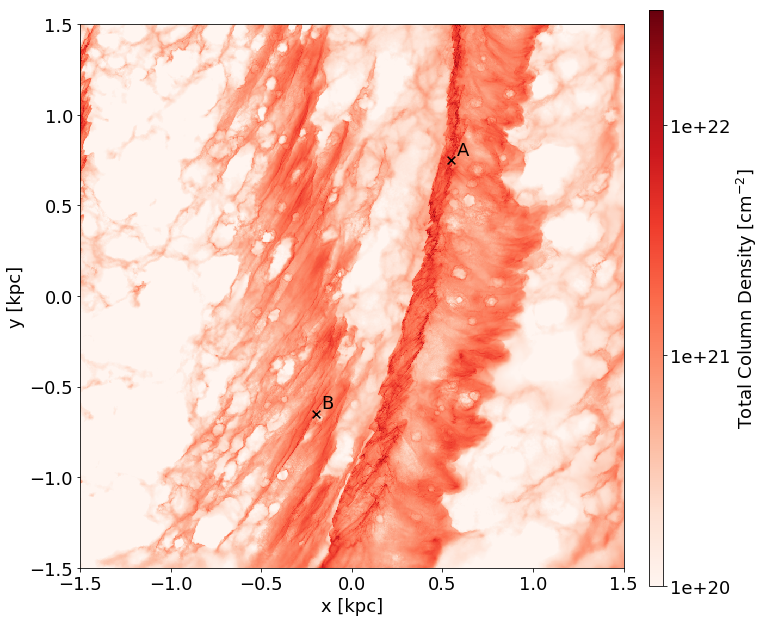}
\\ 
\begin{overpic}[scale=0.32]{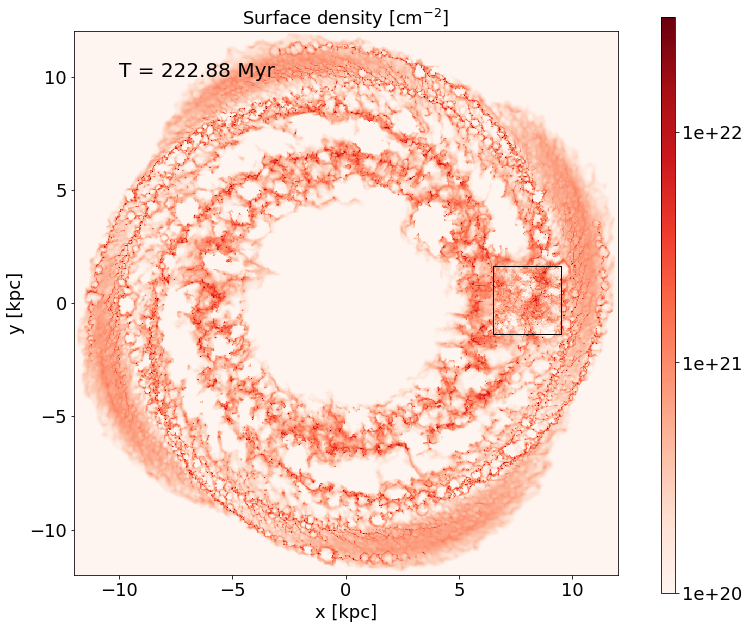}
\put (30,10) {\makebox(0,0){{\bf Feedback dominated}}}
\end{overpic}
\includegraphics[width=\columnwidth]{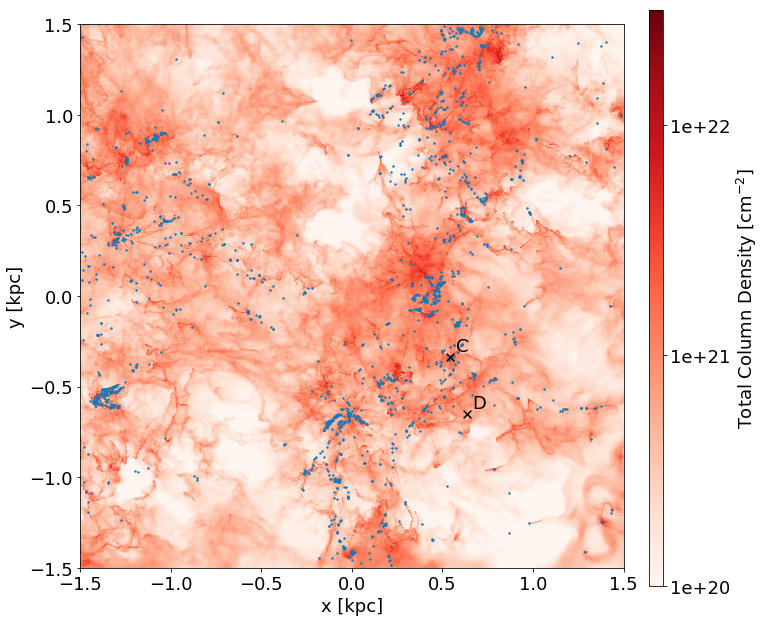}

\end{tabular}
\caption{A face on view of the galactic-scale gas distribution in the potential and feedback dominated cases. The black box shows the location of the 10 \msun\ resolution region, which is shown in more detail in the right panel. Blue dots show the locations of sink particles. Letters show the location of cloud complexes A, B, C and D.}
\label{fig:galaxyview}
\end{center}
\end{figure*}

\subsection{The \textsc{arepo} code}
\label{sec:code}
We perform our simulations using a version of the \textsc{arepo} code \citep{Springel10,Pakmor16} modified to include our custom ISM physics modules. \textsc{arepo} is a well tested cosmological code that solves the (M)HD equations on a Voronoi mesh (for this study we do not include magnetic fields - but full MHD runs will be presented in future work). This mesh is adaptable and can be refined to give improved mass resolution in regions of interest, and the time-stepping is variable, making it the ideal tool for problems with a large dynamic range like star formation in galaxies. We have modified the base \textsc{arepo} code to include the following features to model the cold interstellar medium.

\subsection{Galactic potential}
\label{sec:potential}

The aim of the simulations presented here is not to simulate the self-consistent evolution of a spiral galaxy, but rather to study how the ISM responds to large-scale galactic effects such as differential rotation, spiral arms, and supernova feedback bubbles. We therefore chose to model the large-scale galactic potential analytically in order to reduce computational effort and make a clean controlled test. The self-gravity of the gas itself is calculated using the standard \arepo gravitational tree \citep{Springel10}. For the axisymmetric part of our analytic gravitational potential, we use the best fitting potential of \cite{McMillan17}, which was created to be consistent with various observational and theoretical constraints for the Milky Way. It is the sum of a bulge, disc, and halo component, which are assumed to be generated by the following density distributions:

\subparagraph{Bulge:} This component is generated by the following density distribution:
\begin{equation}
\rho_{\rm b} = \frac{ \rho_{{\rm b}0} }{(1 + a/a_0)^\alpha} \exp\left[ - \left( a/a_{\rm cut}\right)^2 \right]
\end{equation}
where
\begin{equation}
a = \sqrt{x^2 + y^2 + \frac{z^2}{q_{\rm b}^2}},
\end{equation}
and $\alpha=1.8$, $a_0=0.075\kpc$, $a_{\rm cut}=1.9\kpc$, $q_{\rm b}=0.5$ and $\rho_{\rm b0}=9.93\times10^{10}\, \rm M_\odot \kpc^{-3}$.

\subparagraph{Disc:} We assume that the disc is the sum of a thick and a thin disc \citep{Gilmore83}. The density distribution is:
\begin{equation}
\rho_{\rm d} = \frac{\Sigma_1}{2 z_1} \exp \left( -\frac{|z|}{z_1} - \frac{R}{R_{{\rm d} 1}} \right) + \frac{\Sigma_2}{2 z_2} \exp \left( -\frac{|z|}{z_2} - \frac{R}{R_{{\rm d} 2}} \right),
\end{equation}
where $R=\sqrt{x^2+y^2}$ is the cylindrical radius, $\Sigma_1~=~896 {\rm M}_\odot \kpc^{-2}$, $R_{\rm{d}1} = 2.5\kpc$, $z_1=0.3\kpc$, $\Sigma_2~=~183 {\rm M}_\odot \kpc^{-2}$, $R_{\rm{d}2}=3.02\kpc$, and $z_2=0.9\kpc$.

\subparagraph{Halo:} This is a simple \cite{Navarro96} profile. The density distribution is:
\begin{equation}
\rho_{\rm h} = \frac{\rho_{{\rm h}0}}{x (1 + x)^2}
\end{equation}
where $x = r/r_{\rm h}$, $r=\sqrt{x^2+y^2+z^2}$ is the spherical radius, $\rho_{\rm{h}0}=0.00854 M_\odot \pc^{-3}$, and $r_{\rm h} = 19.6\kpc$.

In additional to this axisymmetric potential, we also include a spiral perturbation to the potential, generated in the same way as in \citet{Smith14a}. Briefly, we use a four-armed spiral component from \citet{Cox02} with a pitch angle $\alpha = 15^\circ$ and a pattern speed of $2 \times 10^{-8} \, {\rm rad \, yr^{-1}}$.

\subsection{Gas chemistry and cooling}
\label{sec:chem}
The chemical evolution of the gas is modelled as in \citet{Smith14a} using the hydrogen chemistry of \citet{Glover07a,Glover07b}, together with the highly simplified treatment of CO formation and destruction introduced in \citet{Nelson97}. Our modelling of the hydrogen chemistry includes H$_2$ formation on grains, H$_{2}$ destruction by photo-dissociation, collisional dissociation of atomic hydrogen, H$^{+}$ recombination in the gas phase and on grain surfaces (see Table 1 of \citealt{Glover07a}), and cosmic ray ionisation. The evolution of the CO abundance is calculated assuming that the CO formation rate is limited by an initial radiative association step, and that the CO destruction rate is primarily due to photodissociation. Full details of the combined network, and a discussion of how it compares to other approaches in terms of accuracy and speed, are given in \citet{Glover12a}. The network we use here is the same as the NL97 model in that paper.

We assume that the strength and spectral shape of the ultraviolet portion of the interstellar radiation field (ISRF) are the same as the values for the solar neighbourhood derived by \citet{Draine78} (equivalent to 1.7 times the field strength derived by \citealt{Habing68}). To treat the attenuation of the ISRF due to H$_{2}$ self-shielding, CO self-shielding, the shielding of CO by H$_{2}$, and by dust absorption, we use the \textsc{TreeCol} algorithm developed by \citet{Clark12b} assuming a shielding length of $L_{\rm sh} = 30 \: {\rm pc}$, which roughly corresponds to the to typical distance to the nearest O or B star in the solar neighbourhood \citep{Reed00}. Our field represents the general background radiation field and does not explicitly include ionising radiation from massive stars. We discuss this caveat further in Section \ref{sec:future}. Heating and cooling of the gas from radiative processes is computed alongside the chemistry using the atomic and molecular cooling function described in \citet{Clark19}. In this latest version of our cooling function, the cooling of high temperature gas ($T>10^4$~K) via atomic hydrogen line emission is modelled using H and e$^{-}$ abundances taken directly from our non-equilibrium chemical model. High temperature cooling from helium and metals, on the other hand, is computed assuming that these are in collisional ionisation equilibrium, using values taken from \citet{Gnat12}. We adopt a cosmic ray ionisation rate of $\xi_{\rm H} = 3 \times 10^{-17}$ s$^{-1}$ for atomic hydrogen, and a rate twice this for molecular hydrogen. Finally we assume a solar metal abundance, and a 100:1 gas-to-dust ratio.

\subsection{Modelling star formation}
\label{sec:sinks}
Star particles are commonly used in galactic simulations to represent locations where clusters of stars are formed and feedback will be injected  \citep[e.g.][]{Dobbs10,Hopkins12a,Schaye15}, whereas in simulations of molecular clouds sink particles, typically representing individual stellar systems, are used \citep[e.g.][]{Bate95,Federrath10a}. Due to our varying resolution we must adopt a hybrid approach. 

We use the framework of sink particles as our base. Sink particles are non-gaseous particles that represent sites of star formation. Cells with densities exceeding a critical density, $\rho_{c}$, are candidates for conversion to sink particles, but must first successfully pass a series of energy checks that verify whether the gas is unambiguously gravitationally bound and has inwardly directed velocities and accelerations. In addition to these checks, before a cell is transformed to a sink it must be located at a local minimum in the gravitational potential, and outside of the accretion radius of any existing sink particle. Further details of our sink particle creation algorithm can be found in \citet{Tress19}. As we will use varying levels of resolution in our simulations (see Sections \ref{sec:ics} and \ref{sec:zooms}), we also have varying critical sink creation densities depending on the target mass resolution. These are described in Table \ref{tab:sinkprops}. Even at our lowest creation density the gas has temperatures of order 40 K and so we are still modelling the formation of cold gas. At all points in the simulation we require that the Jeans length is resolved by at least 4 cells \citep{Truelove97} and so even if we exceed these creation densities, because the gas does not pass the energy checks, we continue to resolve the gas and avoid artificial fragmentation (see \citealt{Greif11}).

\begin{table}
	\centering
	\caption{Sink properties as a function of target resolution mass. $\rho_{\rm c}$ is the sink creation density }
		\begin{tabular}{c c}
		\hline
	         \hline
	          Target Mass [\msun] & $\rho_{\rm c}$ [\cmc]  \\
	   	 \hline
		 $\geq$ 100. & 100.\\ 
		 10. & 574.\\ 
		 0.25 & 10000. \\ 
	         \hline	 
	         \hline        
		\end{tabular}
	\label{tab:sinkprops}
\end{table}

The bound gas is replaced with a sink particle of a given accretion radius which will accrete from neighbouring bound cells by skimming mass from them. We employ a variable sink accretion radius in this study. Initially the sinks form with an accretion radius that is chosen to match the Jeans length at their creation density, assuming a temperature of 10K. The accretion radius then grows in time such that the `density'. in the sink, $\rho_{\rm sink} = \frac{3\, M_{\rm sink}}{4\, \pi \, R_{\rm acc}^3}$, remains constant in time. This has the effect that the acceleration at the sink surface remains constant with time, effectively setting a (rough) lower bound to the time-step hierarchy in the simulation. The exception to this is in our tracer refinement regions (see Section~\ref{sec:zooms} below), where we set a constant accretion radius of a parsec as we want to focus on the filament properties and not the protostellar core masses in this first paper.

If a cell denser than $\rho_{\rm c}$ comes within $r_{\text{acc}}$ of a sink cell we check if it is gravitationally bound to the sink, and if so transfer an amount $\Delta m = (\rho_{\text{cell}} - \rho_{\rm c}) V_{\text{cell}}$ of mass from the cell to the sink, where $\rho_{\text{cell}}$ and $V_{\text{cell}}$ are the accreted cell's density and volume. For stability, $\Delta m$ is limited to a maximum of 90\% of the initial mass of the cell. If a cell is within the accretion radii of multiple sinks then the mass is transferred to the sink to which it is most bound. As the sinks represent small clusters rather than point masses, we set their gravitational softening radius equal to the accretion radius to avoid artificially large gravitational accelerations. As \arepo has hierarchical timesteps, we require that sinks are evolved on the shortest timestep of any gas cell in the simulation. Without this restriction, we will potentially miss accretion from cells that spend only a short time within the sink accretion radius if they happen to have moved outside of $r_{\rm acc}$ by the beginning of the next sink timestep.

As the sinks are formed at densities below those of star-forming cores (particularly for the high target mass cells) we assume that not all the mass is converted into stars. Molecular clouds are observed to have low star formation efficiencies \citep[see e.g.][]{Krumholz07} of just a few percent. Based on this work we use a rough star formation efficiency of 1-2 \% for the large target masses. For our highest resolution regions formed at higher densities we use a correspondingly higher star formation efficiency of $33\%$ \citep{Matzner00}. We then multiply the mass of the sink by the assumed star formation efficiency to get the stellar content of the sinks. To calculate the number of massive stars that contribute to this stellar mass we use the approach of \citet{Sormani17} where a stellar initial mass function (IMF; here we use \citealt{Kroupa02}) is binned in mass and the number of stars in each bin is chosen according to Poisson sampling with a mean appropriate to the chosen IMF. Due to the additive properties of Poisson statistics this means that it does not matter if the sinks are large or small, or if matter is accreted after the sink is formed, as the average distribution of stellar masses will be the same everywhere.

Ultimately, our sink particles can represent everything from multiple systems up to large clusters depending on the resolution. However, in practice we will only \textit{analyse} regions where the sinks represent small clusters formed from a single collapsing gas clump. The largest sinks are used simply to track the mass involved in star formation in order to set the feedback returned to the disc as described in the next Section. 

\subsection{Supernova feedback}
\label{sec:feedback}

Supernovae are one of the most important sources of feedback in the ISM, injecting not just thermal energy but also momentum into the surrounding gas. To track the appropriate feedback rate we use two approaches: 1) purely random supernova explosions, and 2) supernovae tied to sinks. In the first case we randomly sample points from the initial gas density profile chosen for the disc (see Section \ref{sec:ics}) with an assumed rate of 1 supernova per 50 years, which is typical of the Milky Way \citep{Diehl06}. However, it is known that purely random feedback can give unrealistic cloud properties when self-gravity is included as it cannot destroy large molecular cloud complexes \citep{Gatto15,Walch15}. Our second feedback approach addresses this by using both a randomly distributed supernova component of 1 supernova every 300 years \citep{Tsujimoto95} to represent Type Ia supernovae, but also including supernovae from sink particles. This is done for all sink particles regardless of the level of refinement that they were formed at.  

For each massive star greater than 8 \msun \ associated with a given sink we trigger a supernova explosion at the end of its lifetime (taken from \citealt{Maeder09}) and inject $10^{51}$ erg of energy into its surrounding gas. Either thermal energy or momentum is injected into the gas depending on whether the Sedov-Taylor phase of the expansion is resolved. We calculate the Sedov-Taylor radius $R_{\rm ST}$ using the mean density of the injection region following the approach of \citet{Gatto15}, so that 
\begin{equation}
R_{\rm ST}=19.1 \left(\frac{E_{\rm SN}}{10^{51} {\rm erg}}\right)^{5/17}\left(\frac{\bar{n}}{{\rm cm}^{-3}}\right)^{-7/17} \, {\rm pc},
\end{equation}
where $E_{\rm SN}$ is the energy of the supernova and $\bar{n}$ is the mean number density within the injection region. We require each supernova injection region to contain at least 32 cells, and so if the injection radius given by this requirement is less than $R_{\rm ST}$ the blast wave phase is resolved and we can inject thermal energy directly. However, if the injection radius is larger than $R_{\rm ST}$ then this phase is unresolved and we instead inject the terminal momentum $p_{\rm ST}$ into the cells following \citet{Blondin98}:
\begin{equation}
p_{\rm ST} = 2.6 \times 10^5 \left(\frac{E_{\rm SN}}{10^{51} {\rm erg}}\right)^{16/17} \left(\frac{\bar{n}}{{\rm cm}^{-3}}\right)^{-2/17} \textrm{M$_\odot$ km s$^{-1}$}.
\end{equation}

As the supernovae from each sink explode individually this naturally results in clusters of supernova explosions. To account for the fact that the sinks and thus the star clusters they represent have a finite size we randomly sample where the supernovae will occur using a Gaussian distribution with a width that is twice the accretion radius. During each supernova we return mass from the sink to the ISM to represent gas that is unbound from the star forming region by feedback. Each supernova returns an ejection mass of $M_\text{ej} = (M_\text{sink}-M_\text{stars})/n_{\rm SN}$, where $M_\text{sink}$ is the mass of the sink at the point the supernova occurs, $M_\text{stars}$ is the mass of stars within the sink at that time, and $n_{\rm SN}$ is the remaining number of supernovae scheduled to go off from the sink. For more detail on our supernovae model see \citet{Tress19}.

Of course, supernovae are not the sole source of stellar feedback in the ISM. In reality there will also be contributions from stellar winds, jets and photoionisation regions, and this is an important caveat for this work. In this paper we focus only on the supernovae, due to the difficulty of including all these effects simultaneously in a galactic-scale simulation. However, we aim to return to this in future work.

\subsection{Simulation setup and refinement}
\label{sec:ics}

We begin our simulations by setting up a gas disc inspired by the Milky Way gas disc model of \citet{McMillan16} that is based on a combination of observational constraints and theoretical modelling. This consists of two density distributions for the HI and H$_2$ that decline exponentially at large radii. Since we start our simulations from an atomic state we add both these contributions together for our initial condition (molecular hydrogen will soon form self-consistently as the gas disc evolves). As we focus on Milky Way like clouds outside the central bar we neglect galactic radii smaller than 4 kpc and greater than 12 kpc for reasons of computational efficiency (for an investigation of the Galactic centre using our modified version of \arepo, see \citealt{Sormani18}). The gas disc is given the initial rotation curve that arises from the galactic potential described in Section \ref{sec:potential}, which for our disc corresponds to a rotation curve of order 220 \kms.

For the first 150 Myr of the simulation we simply let the gas distribution respond to the large-scale potential and develop spiral arms with purely random supernova feedback and no gas self-gravity. During this period we set the refinement such that each cell has a target mass resolution of 1000 \msun. After 150 Myr, when we have reached a steady state, we begin the middle phase (the final phase will be discussed in Section \ref{sec:zooms}) of the simulation by turning on refinement for two spiral arm passages ($\sim 70$ Myr) within a 3 kpc box that co-rotates with the gas centred on a galactic radius of 8 kpc. In this high resolution region the gas has a target mass of initially 100 \msun for the first 60 Myr, but it is further lowered to 10 \msun \ for the final 10 Myr. As previously mentioned we additionally always require that the Jeans length is resolved by at least four cells up to our sink creation density. To avoid discontinuous jumps in the cell size, particularly where the target resolution is changing at the boundaries of the high resolution box, we require that the cell radius of adjacent \textsc{Arepo} cells can differ by no more than a factor of two at any time throughout the entire simulation volume.

For the purposes of this work we run two different versions of this middle phase. In the potential-dominated case gas self-gravity remains turned off, there is no sink formation, and the supernova feedback is purely random, so the gas dynamics is mainly determined by the large-scale gravitational potential. In the feedback-dominated case we turn gas self-gravity on, allow sink particles to form, and use the mixed feedback injection scheme with supernovae tied to the sink particles as described in Section \ref{sec:feedback}. This results in a strong burst of feedback after the self-gravity is turned on that disrupts the gas in the spiral arms. The differing physical forces applied to the two high simulations are shown in Figure \ref{fig:galaxyview} are summarised in Table \ref{tab:box_physics}.

 \begin{table*}
	\centering
	\caption{The target resolution and differing physics applied in the two simulations of the high resolution 3 kpc sized boxes shown in Figure \ref{fig:galaxyview}.}
		\begin{tabular}{l c c c c}
		\hline
	         \hline
	          Simulation & Highest target resolution in box & Feedback & Gas self-gravity & Sinks  \\
	   	 \hline
		 Potential Dominated & 10 \msun & Random & Off & No \\ 
		 Feedback Dominated & 10 \msun & Mixed & On & Yes \\ 
	         \hline	 
	         \hline        
		\end{tabular}
	\label{tab:box_physics}
\end{table*}

Figure \ref{fig:galaxyview} shows the column density of the simulations at the end of the middle phase. The top panels show the first case without self-gravity and with random feedback. In this case the random supernovae are inefficient at pushing around the dense gas and so the large-scale potential dominates the dynamics leading to clear sharp spiral arms. The bottom panel shows the case with self-gravity and mixed feedback. Here the feedback is far more effective at disrupting the gas and the distribution is more irregular and tenuous. The left panels show the overall view of the galactic disc and a box shows the location of the 10 \msun \ solar mass resolution region that co-rotates with the gas. The right panels show this 3 kpc region in more detail. Blue dots show the location of sink particles in the feedback-dominated case. These largely follow the outline of the spiral arms in the above panel. These set-ups are deliberately designed to be extreme in order to allow us to make a clean comparison between the very large scale effects of the galaxy potential and differential rotation, and the more local effects of supernova bubbles on the resulting cloud properties.

\subsection{The ISM on large scales}
\label{sec:ISM}

\begin{figure}
	\includegraphics[width=7cm]{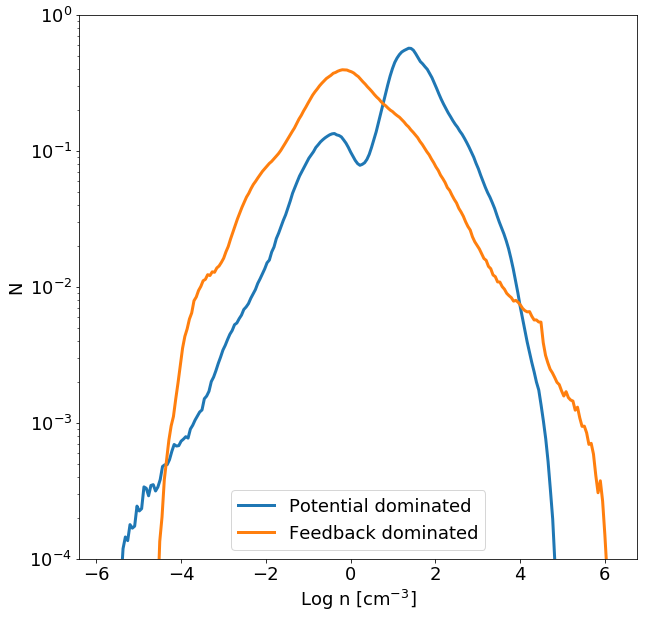}
    \caption{Mass-weighted PDF of the density distribution of the high resolution boxes in the potential-dominated and feedback-dominated simulations. Note that these should not be directly compared as in the feedback-dominated case dense gas may be inside sink particles.}
   \label{fig:pdf_box}
\end{figure}

Figure \ref{fig:pdf_box} shows the mass-weighted density PDF of the \arepo gas cells in the 10 \msun\ resolution box in the two cases. Higher gas densities are reached in the feedback-dominated case as it includes self-gravity. However, there is less dense ($n>100$ \cmc) gas in total due to (i) gas being locked up in sink particles, and (ii) the increased feedback meaning there is more supportive turbulence in the gas.

\begin{figure*}
\begin{center}
\begin{tabular}{c c}
\begin{overpic}[scale=0.3]{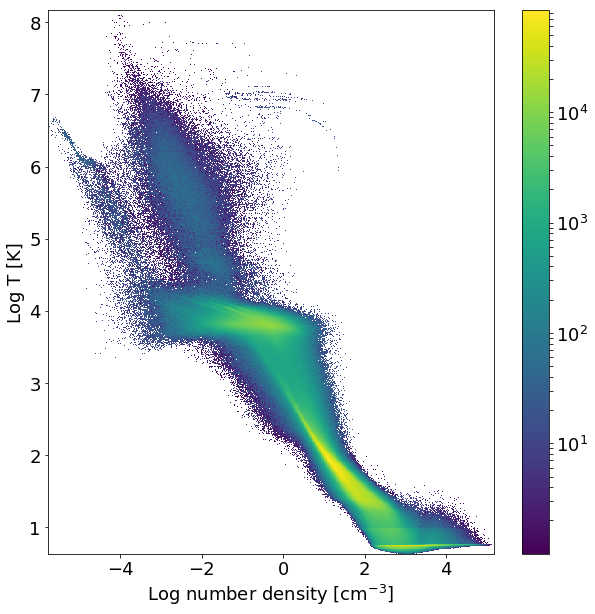}
\put (60,90) {\makebox(0,0){{\bf Potential dominated}}}
\end{overpic}

\begin{overpic}[scale=0.3]{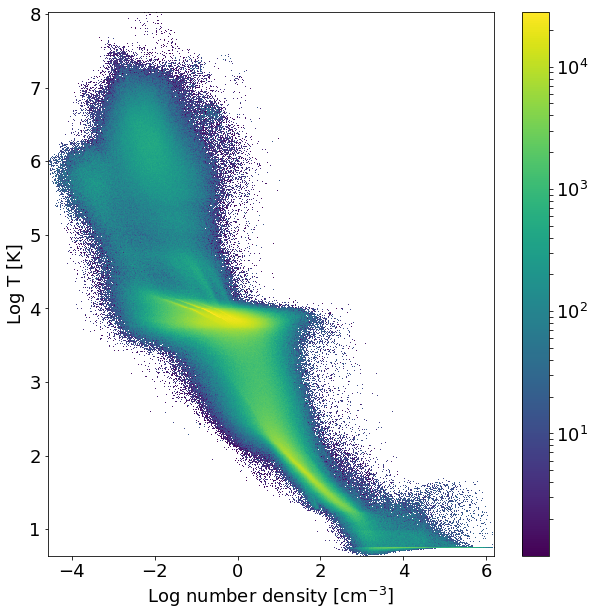}
\put (60,90) {\makebox(0,0){{\bf Feedback dominated}}}
\end{overpic} \\

\begin{overpic}[scale=0.25]{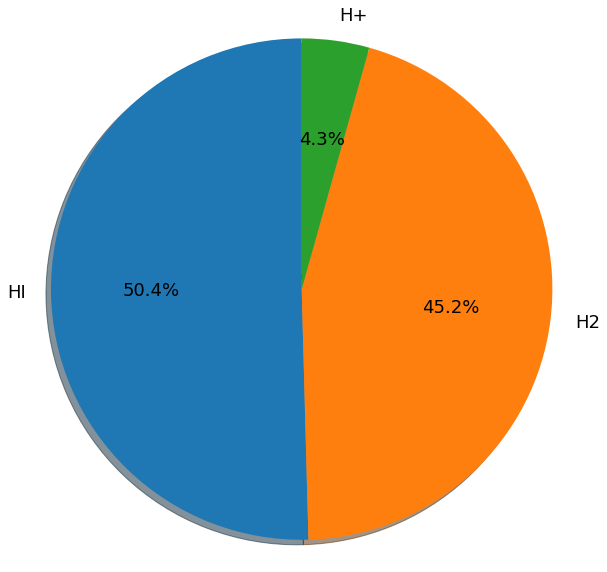}
\end{overpic}

\begin{overpic}[scale=0.25]{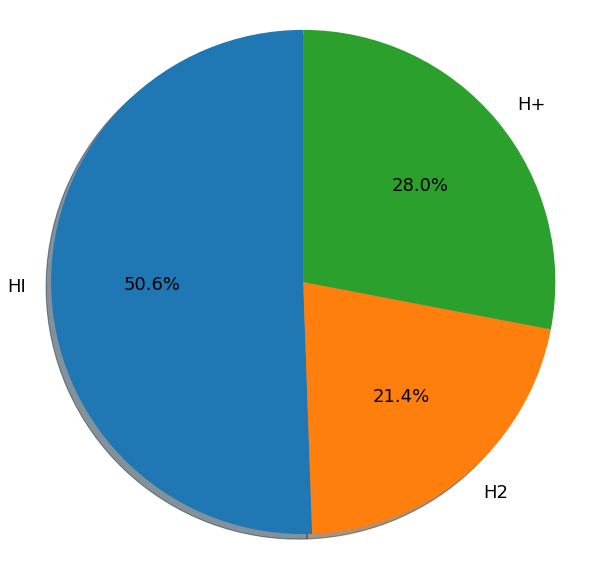}
\end{overpic} \\

\end{tabular}
\caption{The thermodynamic and chemical state of the gas in the 3 kpc, 10 \msun\  resolution region at the time shown in Figure \ref{fig:galaxyview}. The top panels show a mass-weighted 2D histogram of the thermodynamic state of the gas cells where the colour scale shows the counts per bin. The  The left panels show the potential-dominated case, in which due to the lack of strong feedback the gas has a high fraction of cold molecular gas. The right panels show the feedback-dominated case where there is now a substantial amount of warm ionised gas. For the feedback case the H$_2$ mass in the bottom panels includes the mass in young sinks, which represent young star-forming clouds that are likely to still have substantial molecular mass.}
\label{fig:Box_gas_phase}
\end{center}
\end{figure*}

Figure \ref{fig:Box_gas_phase} shows the thermodynamic and chemical state of the gas. The top panels shows a mass-weighted 2D histogram of the phase space distribution of number density vs temperature. Without clustered supernova feedback driving large bubbles there is more cold dense gas in the potential-dominated case. Likewise, when considering the chemical makeup of the gas, a substantially higher fraction is molecular hydrogen in the potential-dominated case (45\% vs 21\%). As the feedback-dominated case has dense gas locked up in sink particles we include in the molecular hydrogen total the mass of young sinks formed within the last 4 Myr, which we interpret as star-forming molecular clouds where feedback will not yet have substantially disrupted the dense molecular gas (without including this mass the molecular total would fall to 11.3\%). We make no comment on the CO fraction of the ISM within the boxes as this will not be fully converged at these resolutions \citep{Seifried17}, but will return to this later.

\subsection{Enhanced resolution using tracer-based refinement}
\label{sec:zooms}

In the final stage of the calculation, after we have generated a diverse variety of cold ISM structures at 10 \msun \ resolution, we further refine individual clouds by injecting massless Monte Carlo tracer particles into regions of interest which are advected probabilistically with the gas flow (see \citet{Genel13} for more details). We select regions of interest from both the potential-dominated and feedback-dominated simulations. In the potential-dominated case, self-gravity and sink formation are turned on once the tracers are injected so that we may study the star formation properties of the gas. Our regions of interest are chosen to represent four contrasting scenarios, described further below.

Within each of these four regions, tracers are injected within a 100 pc radius region everywhere the gas density is above 100 \cmc, with 40 tracer particles being injected per solar mass to ensure that tracers will be present in every relevant \arepo cell even at high refinements. As the clouds evolve the tracer particles move with the gas and can be used as tags to label cells that should be refined to even higher resolutions. In this way we can resolve substructures within individual cloud complexes without neglecting large-scale effects outside the cloud.

\begin{figure}
	\includegraphics[width=7cm]{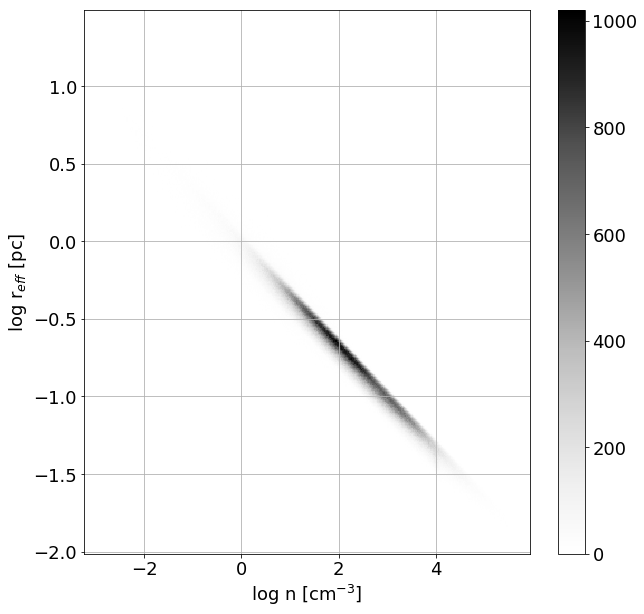}
    \caption{A 2D histogram showing the resolution of the \arepo cells
    in terms of their effective radius as a function of number density in the tracer particle region. The grey-scale shows the number of cells. At number densities of $10^3$ \cmc and above we have a cell radius of 0.1 pc or better meaning that the clouds are well-resolved.}
   \label{fig:resolution}
\end{figure}

Where tracer particles are present we further increase the resolution target mass to 0.25 \msun and again require that the Jeans length is resolved with a minimum of 4 \arepo cells until number densities of $n=10^4$ \cmc. Figure \ref{fig:resolution} shows the resulting spatial resolution that arises from this requirement in a typical tracer refinement region. At number densities of $10^3$ \cmc and above we have a cell radius of 0.1 pc or better.

 Four regions from the two cases are selected as illustrative examples of clouds experiencing different conditions for subsequent tracer refinement. We will refer to these as cloud complexes, as they have a complex geometry and contain several smaller molecular clouds. In the potential dominated case with purely random feedback we investigate the behaviour of gas inside and outside the spiral arm. Complex A is chosen to be inside an arm, and B is an inter-arm region. In the mixed feedback case where large supernovae bubbles disrupt the gas we select the two highest density regions where there are not already massive sink particles (complexes C and D). The more massive of these, complex D, subsequently forms many sinks leading to additional supernova feedback from within the clouds as it forms new massive stars. 
 
We let the tracer refinement regions evolve for at least another 4 Myr in total. Table \ref{tab:regions} summarises the total mass of cells containing tracer particles at 1 Myr after they were first injected. Note that the mass in the mixed supernova regions is lower as the previous feedback has reduced the density of massive cloud complexes. The total mass in the disc is the same in both cases, but with mixed feedback gas is less concentrated in dense clouds. One new supernova detonates in complex A within the studied period with tracer refinement, and none in complex B. In complex C no new supernovae detonate during the studied tracer-refined period. Complex D has 4 additional supernovae. It should be noted, however, that both complexes C and D originate from an initial condition in which there has been a large burst of clustered supernovae feedback, which has injected energy into the gas.

\begin{table*}[t]
	\centering
	\caption{Summary of the properties of the tracer-refined cloud complexes. The initial condition denotes the high resolution box the tracer particles were injected into. The positions are shown in the right panels of Figure \ref{fig:galaxyview}. The mass is that of the total mass of cells containing tracer particles at 1 Myr in each complex.}
		\begin{tabular}{l c c c c c}
		\hline
	         \hline
	          Cloud Complex & Initial Condition & Feedback & Self-gravity & Refined Mass [\msun] & Description \\
	   	 \hline
		 A & Potential Dominated &  Random & On & $8.00 \times 10^5$ & Spiral Arm \\ 
		 B & Potential Dominated & Random & On & $6.50 \times 10^5$ & Inter-arm \\ 
		 C & Feedback Dominated & Mixed & On & $0.97 \times 10^5$ & Supernova influenced \\
		 D & Feedback Dominated & Mixed & On & $1.74 \times 10^5$ & Embedded supernova \\
	         \hline	 
	         \hline        
		\end{tabular}
	\label{tab:regions}
\end{table*}

As we only analyse the gas in the complexes that have been tracer-refined, it is a valid question to worry if dense gas is later formed within the cloud complexes after the tracers have been injected that will be missing from our analysis. To investigate this we search for gas within an 100 pc radius of the mean position of our tracer particles that is above a number density of 100\cmc but is \textit{not} tracer-refined. We test each complex 2 Myr after the tracer injection time to allow substantial evolution within the cloud. In complexes B, C and D less than 0.5\% of the dense (n>100 \cmc) gas mass fraction is unrefined at 2 Myr after the tracers are injected (B- 0.1\%, C- 0.3\%, D-0.2\%). In these complexes this unrefined dense gas takes the form of only a few cells around the edges of the structure. In complex A, the unrefined dense gas fraction within 100 pc of the mean tracer position is almost 19\%. However, a visual inspection shows that this is not due to dense gas within the clouds being unrefined, but instead to a new dense cloud/region coming within the search radius. This unrefined region can be seen later in Figure \ref{fig:galaxyview} at a position of around $x=-15$ kpc,$y=-65$. As this gas represents a new distinct structure it does not interfere with our analysis of Complex A.

\section{Results}
\label{sec:results}

\subsection{Column density maps}
\label{sec:cl_columns}

\begin{figure*}
\begin{center}
\begin{tabular}{c c}
\begin{overpic}[scale=0.35]{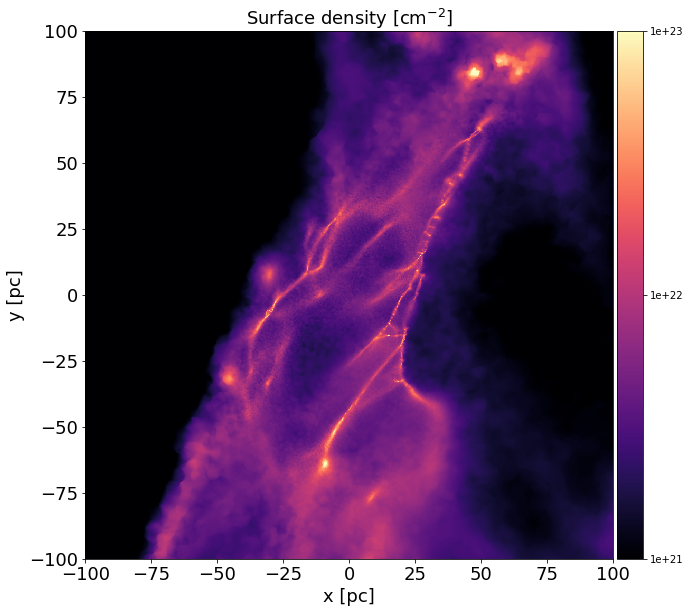}
\put (25,80) {\makebox(0,0){{\bf \textcolor{white}{Complex A}}}}
\put (25,75) {\makebox(0,0){{\bf \textcolor{white}{Spiral Arm}}}}
\end{overpic}

\begin{overpic}[scale=0.35]{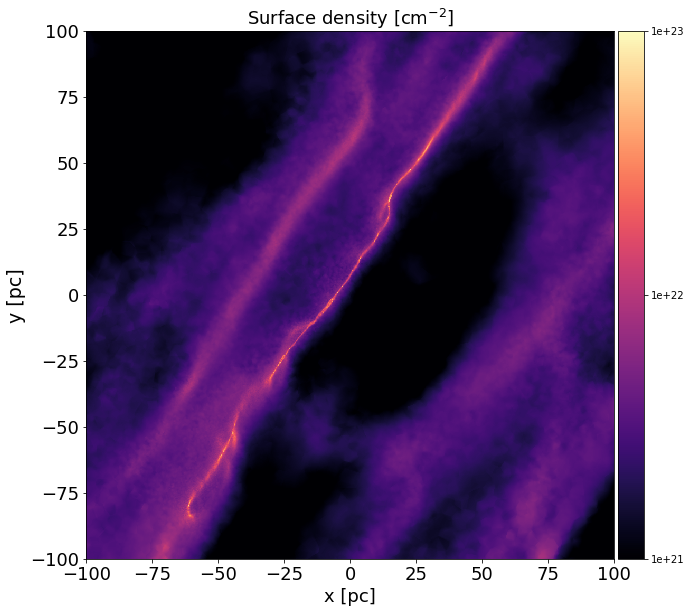}
\put (25,80) {\makebox(0,0){{\bf \textcolor{white}{Complex B}}}}
\put (31,75) {\makebox(0,0){{\bf \textcolor{white}{Inter-Arm Filament}}}}
\end{overpic} \\

\begin{overpic}[scale=0.35]{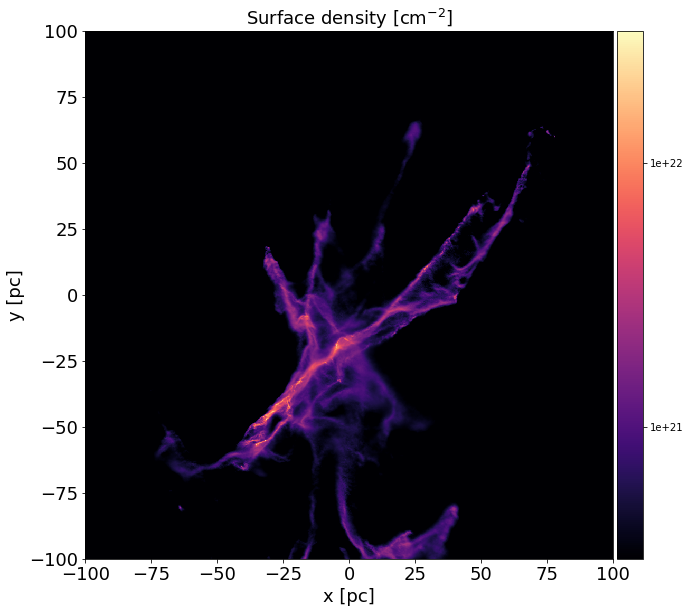}
\put (25,80) {\makebox(0,0){{\bf \textcolor{white}{Complex C}}}}
\put (31,75) {\makebox(0,0){{\bf \textcolor{white}{Supernova influenced}}}}
\end{overpic}

\begin{overpic}[scale=0.35]{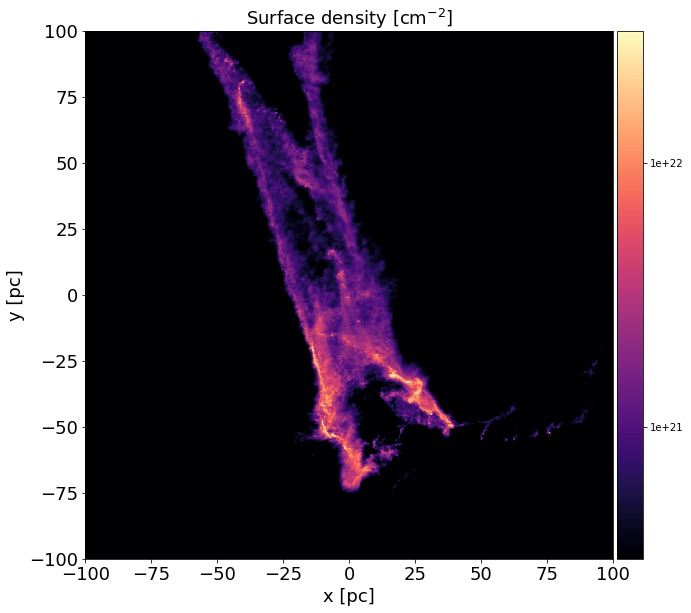}
\put (25,80) {\makebox(0,0){{\bf \textcolor{white}{Complex D}}}}
\put (31,75) {\makebox(0,0){{\bf \textcolor{white}{Embedded supernovae}}}}
\end{overpic} \\

\end{tabular}
\caption{The projected column density of the four cloud complexes 2 Myr after the tracer-particle-based refinement process has commenced. Complexes A and B are drawn from the potential-dominated run, and complexes C and D from the feedback-dominated run shown in Figure \ref{fig:galaxyview}.}
\label{fig:cl_column_maps}
\end{center}
\end{figure*}
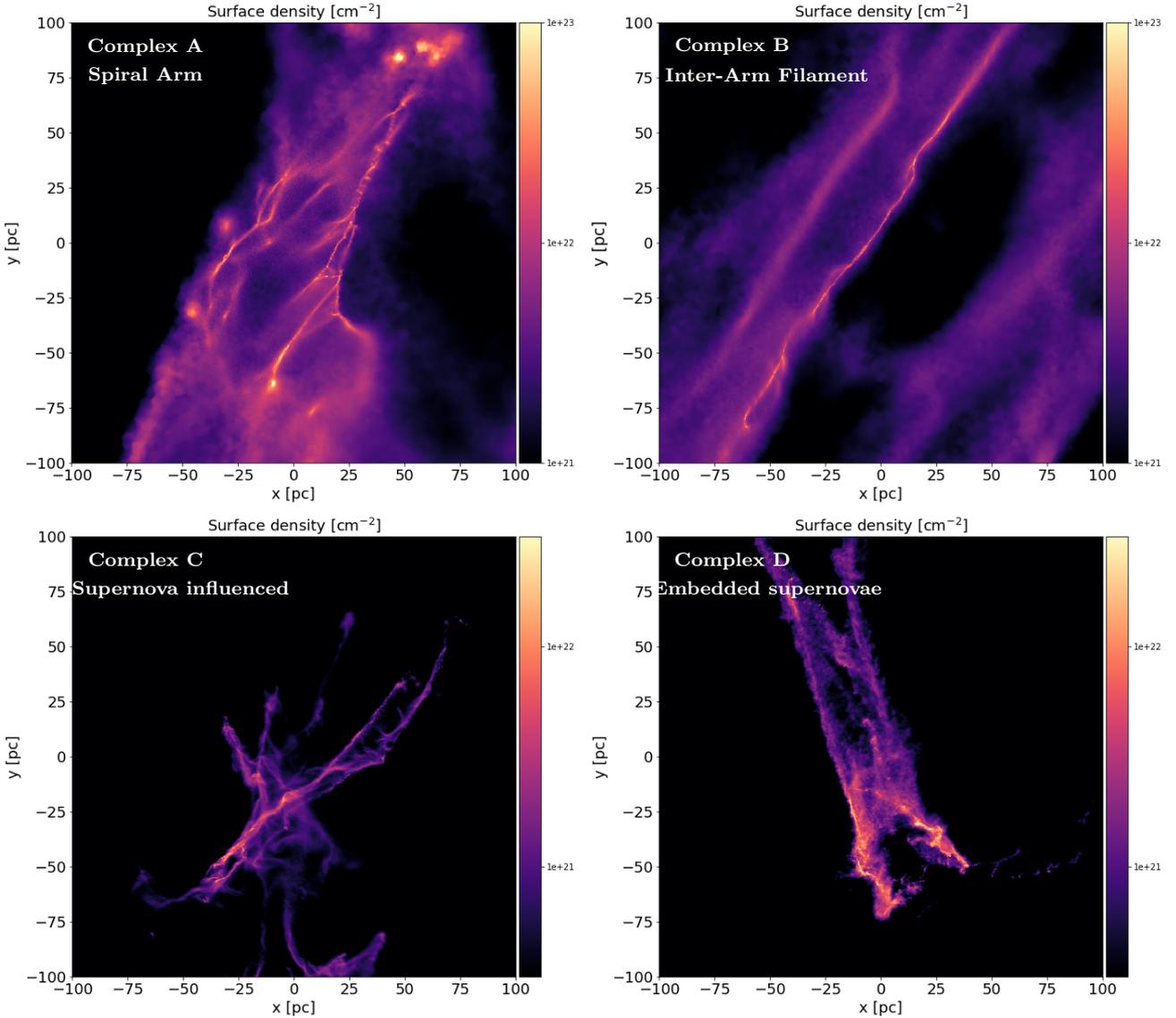

Figure \ref{fig:cl_column_maps} shows the column density maps of the four selected cloud complex regions viewed face-on to the galactic plane. The maps are centred on the centre-of-mass of the cells containing tracer particles, which are the most highly refined regions in the simulation (0.25 \msun \ target mass). Around the outside of the images, particularly in cloud complex A, it can be seen how the highly refined cells smoothly merge into the lower refinement regions. Note how the filamentary features join smoothly between the two, showing that the generation of such structures is not a consequence of the refinement.

Clear differences can be seen in the cloud complex morphologies. When dominated only by the large-scale galactic potential and differential rotation, as in complexes A and B, the clouds have higher column densities and have a smooth continuous filamentary structure. In particular, complex B forms one continuous structure due to a single cloud complex being stretched out by the differential rotation (see also \citealt{Duarte-Cabral17}). However, in the cases where there was clustered feedback and self-gravity prior to the tracer particles being injected, complexes C and D, the column densities are lower and the distribution far more irregular. In the lower-right corner of complex D, we can see that feedback is beginning to disrupt the cloud complex.

\subsection{Gas properties}
\label{sec:gasprops}

\begin{figure}
\begin{center}
\includegraphics[width=\columnwidth]{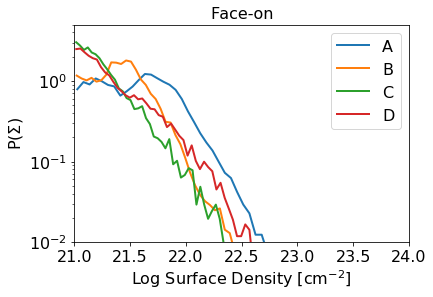}\\
\includegraphics[width=\columnwidth]{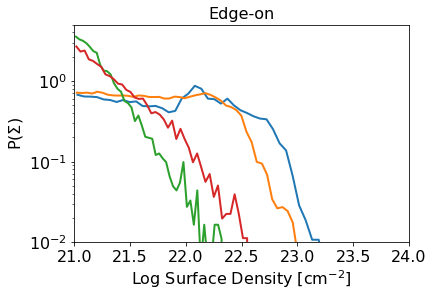}\\
\includegraphics[width=\columnwidth]{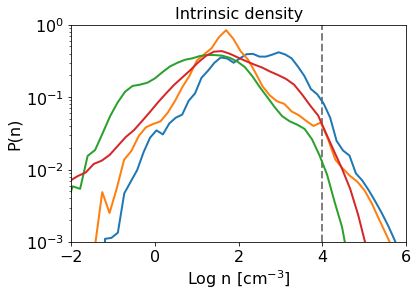}\\
\caption{The probability density distribution of the column densities and intrinsic densities of the gas cells containing tracers. \textit{Top} the projected column density viewed face-on to the galactic plane, \textit{middle} the projected column density viewed edge-on along the y-axis in Figure \ref{fig:cl_column_maps}, and \textit{bottom} the mass-weighted number densities in the clouds. The dashed grey line in the bottom panel shows the critical density for sink creation in the tracer-refined gas.}
\label{fig:complex_pdfs}
\end{center}
\end{figure}

Figure \ref{fig:complex_pdfs} shows the probability density functions (PDFs) of the cloud complex surface densities and volume densities where there are tracer-refined cells. The column density is unsurprisingly higher in the potential-dominated case, especially when viewed edge-on. At lower column densities the PDFs are almost flat for the potential-dominated cases A and B particularly in the edge-on case where due to the gas being tightly confined to the galactic plane there are substantial projection effects. In the feedback dominated cases C and D the PDFs decrease at larger column densities. 

The intrinsic density PDFs of just the tracer-refined \arepo gas cells resemble the lognormal distribution expected for turbulent gas \cite[e.g.][]{MacLow04,Federrath10b}. The density PDFs peak around $n \sim 100$ \cmc for complexes B, C and D, but the peak is shifted to $n \sim 1000$ \cmc for complex A, which is located in a galactic potential dominated spiral arm with low internal turbulence. Given that our sink particles are inserted at number densities of $n \sim 10^4$ \cmc or larger, and that consequently dense gas is missing from the PDF, it is difficult to comment on the existence of any power law tail at high densities that might be evidence of gravitational collapse \citep[e.g.][]{Schneider12}.

\begin{figure}
\begin{center}
\includegraphics[width=\columnwidth]{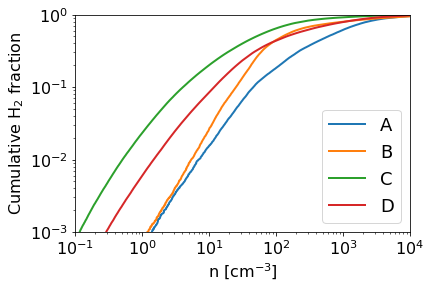}\\
\includegraphics[width=\columnwidth]{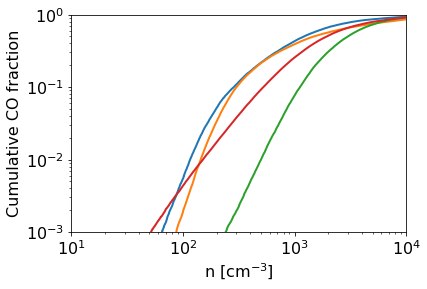}\\
\caption{The cumulative fraction of the H$_2$ \textit{(top)} and CO \textit{(bottom)} mass with increasing number density in each of the cloud complexes. Only gas that has been tracer-refined to our highest resolution level is included in the plots. A greater fraction of the CO mass comes from higher number densities in the feedback dominated complexes C and D.}
\label{fig:gaschemistry}
\end{center}
\end{figure}

In addition to the gas PDFs we can also compare the chemical states of the four cloud complexes. Figure \ref{fig:gaschemistry} shows the cumulative mass of molecular hydrogen and CO with increasing number density in the tracer-refined cells in each of the cloud complexes. There is a greater fraction of the tracer-refined gas in H$_2$ at low number densities in the feedback-dominated complexes C and D compared to the potential-dominated cases A and B. This is probably due to there being more total mass at low densities due to the supernova feedback, but may also reflect mixing of H$_{2}$ from higher density to lower density regions by supernova-driven turbulence \citep{Glover07b,Valdivia16,Seifried17}. Conversely, a greater fraction of the CO in the feedback-dominated cases is at higher densities compared to the potential-dominated cases. This is a consequence of the lower column densities of the cloud complexes in the feedback-dominated run, which make them less effective at shielding CO from the background photo-dissociating radiation field.

\subsection{Filamentary networks}
\label{sec:cl_filprop}

A major motivation for this work is to study the types of filament networks that arise self-consistently within clouds due to differences in the cloud's formation histories. A problem with isolated simulations of cloud formation \cite[e.g][]{Smith16} is that the filamentary structures are in some way pre-determined by the initial conditions as they originated from the initial turbulent velocity field that was prescribed. This is not an issue in the simulations we present here, as the clouds are generated self-consistently from galactic-scale dynamics and feedback.

\begin{figure*}
\begin{center}
\begin{tabular}{c c}
\begin{overpic}[scale=0.35]{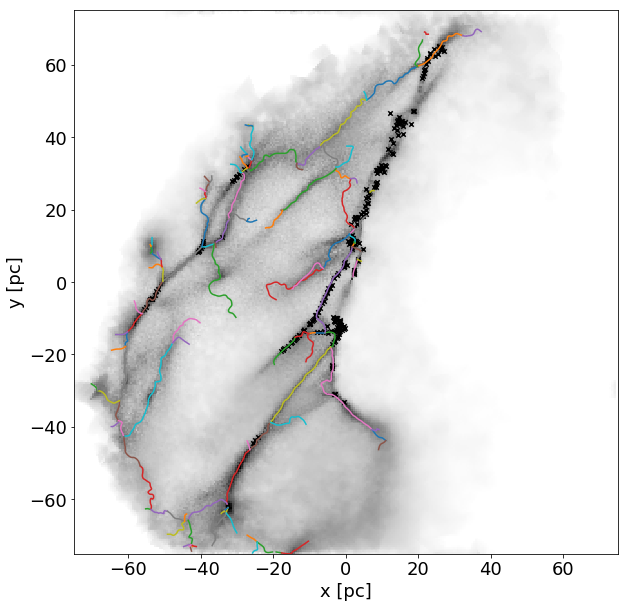}
\put (25,90) {\makebox(0,0){{\bf Complex A}}}
\end{overpic}

\begin{overpic}[scale=0.35]{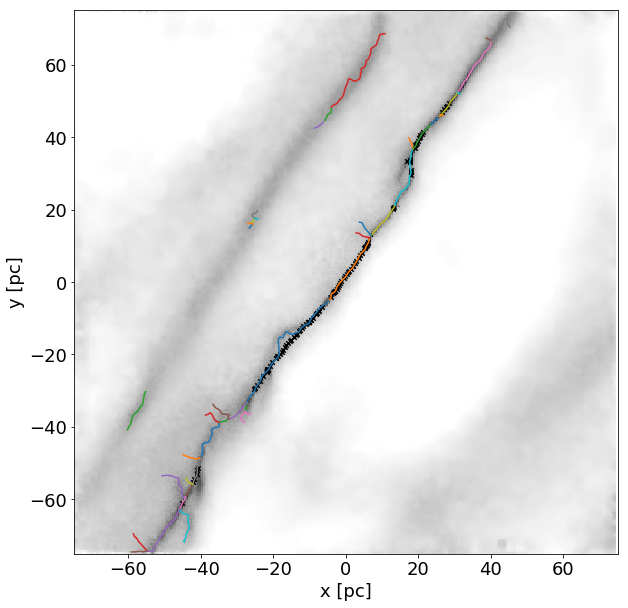}
\put (25,90) {\makebox(0,0){{\bf Complex B}}}
\end{overpic} \\

\begin{overpic}[scale=0.35]{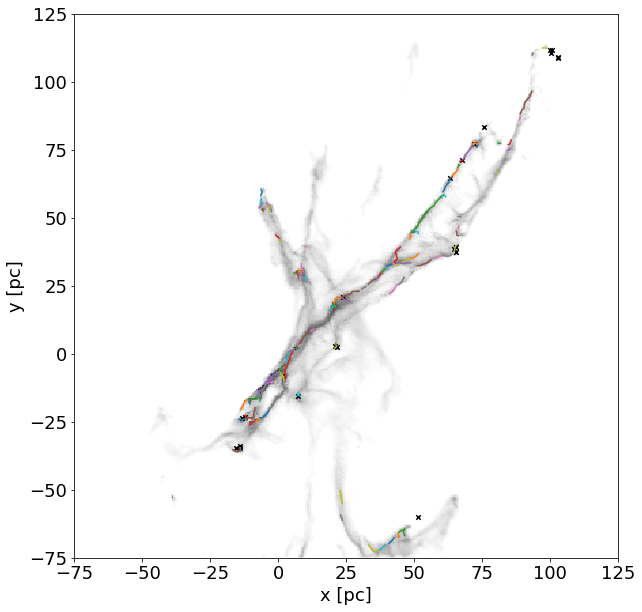}
\put (25,90) {\makebox(0,0){{\bf Complex C}}}
\end{overpic}

\begin{overpic}[scale=0.35]{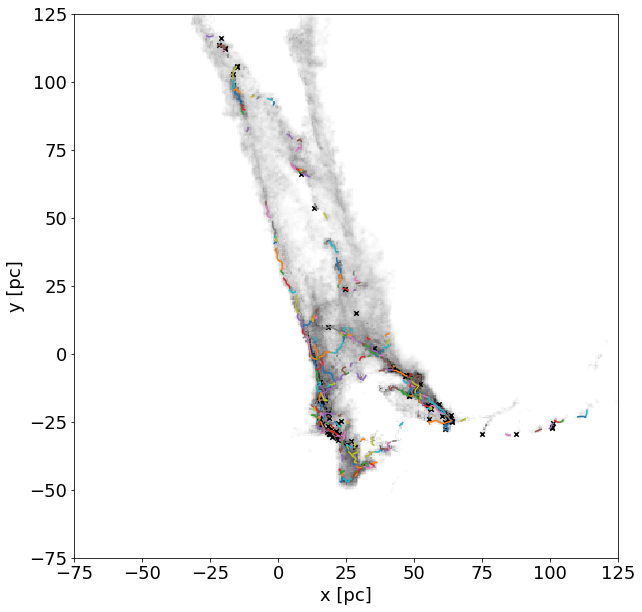}
\put (25,90) {\makebox(0,0){{\bf Complex D}}}
\end{overpic}

\end{tabular}
\caption{The location of all the filament spines identified with \disperse (\textit{coloured lines}) over-plotted on the column density of the gas cells containing tracer particles (\textit{greyscale}) in the $x-y$ plane for the 4 cloud complexes. Black crosses show the location of sink particles representing collapsing cores of star-forming gas.}
\label{fig:fils_xy}
\end{center}
\end{figure*}

\begin{figure*}
\begin{center}
\begin{tabular}{c c}
\begin{overpic}[scale=0.35]{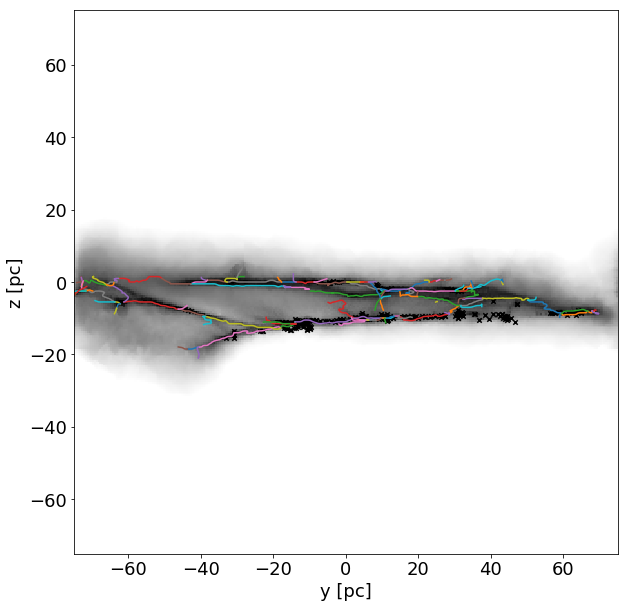}
\put (25,90) {\makebox(0,0){{\bf Complex A}}}
\end{overpic}

\begin{overpic}[scale=0.35]{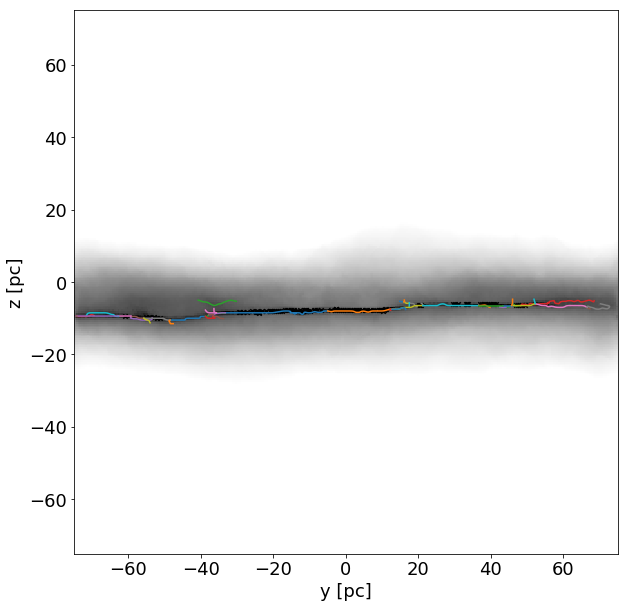}
\put (25,90) {\makebox(0,0){{\bf Complex B}}}
\end{overpic}\\

\begin{overpic}[scale=0.35]{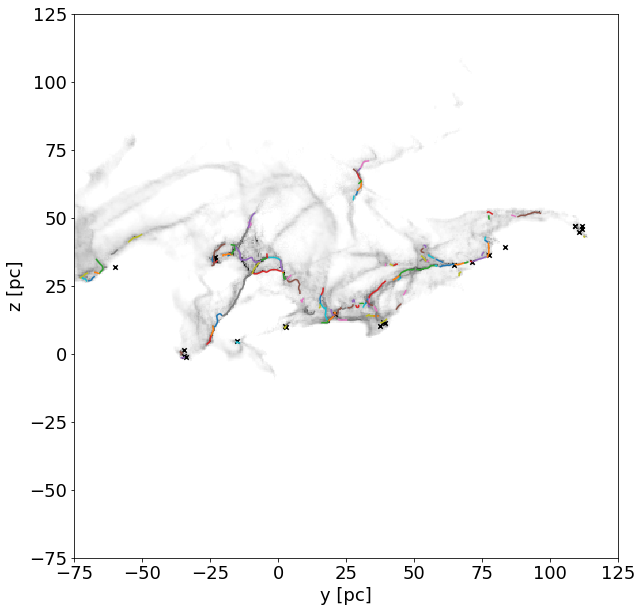}
\put (25,90) {\makebox(0,0){{\bf Complex C}}}
\end{overpic}

\begin{overpic}[scale=0.35]{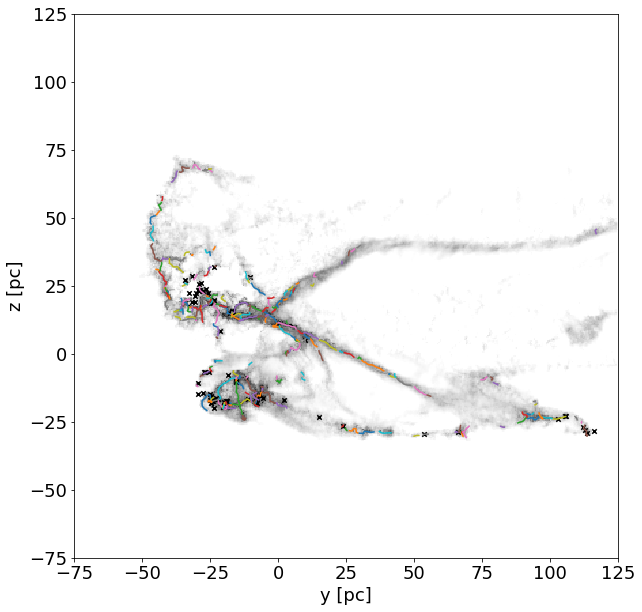}
\put (25,90) {\makebox(0,0){{\bf Complex D}}}
\end{overpic}

\end{tabular}
\caption{The location of filament spines identified with \disperse (\textit{coloured lines}) over-plotted on top of the column density of the gas cells containing tracer particles (\textit{greyscale}) in the $y-z$ plane for the 4 cloud complexes. Black crosses show the location of sink particles representing collapsing cores of star-forming gas.}
\label{fig:fils_yz}
\end{center}
\end{figure*}

To this end we show in Figure \ref{fig:fils_xy} all the filament spines identified using \disperse over-plotted on the column density of the cells with tracer particles in grey scale. Full details of how filaments are identified and gas properties assigned to them is outlined in Appendix \ref{sec:filid}. From this point onward we only include gas cells on our highest level of refinement in the analysis. Black crosses show the location of sink particles, which represent collapsing clumps/cores of star-forming gas. Figure \ref{fig:fils_yz} shows the same gas distribution now viewed within the galactic plane along the x-axis.

As in Figure \ref{fig:cl_column_maps}, immediate differences are apparent between the different clouds. In complexes A and B, where only large-scale galactic forces operated during the cloud's formation before refinement, the filaments identified by \disperse are longer and have vigorously fragmented into star-forming cores along their length. They are almost uniformly parallel to, and tightly confined to the galactic plane. Complexes C and D were formed in an environment with higher turbulence due strong feedback from clustered supernovae. In this case the filaments are shorter and form fewer stars due to the lower mean gas density. The filaments are no longer confined to within 20 pc of the galactic plane and have a range of vertical orientations, approaching perpendicular in some cases.

\begin{figure}
\begin{center}
\includegraphics[width=6.5cm]{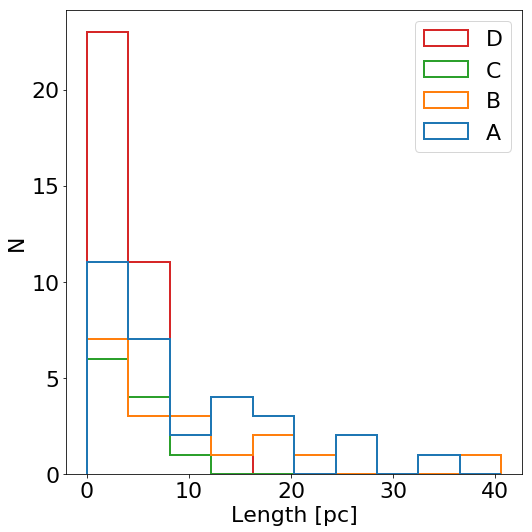}\\
\includegraphics[width=6.5cm]{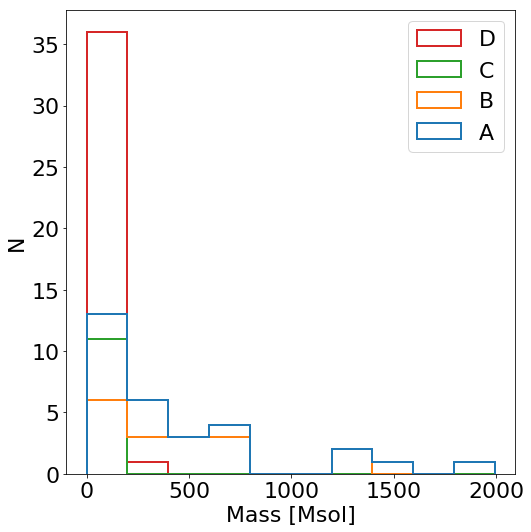}\\
\includegraphics[width=6.5cm]{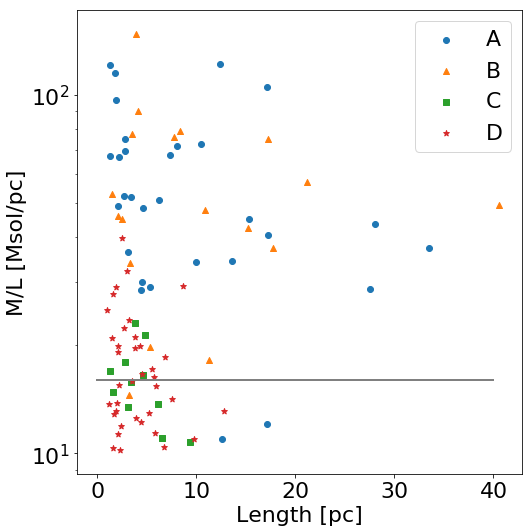}
\caption{The masses and lengths of the resolved filaments identified by \disperse in each cloud complex are shown in the top two panels. The lower panel shows the mass to length ratio for the same filaments, with the grey line showing the critical ratio of 16.7 \msun pc$^{-1}$ for gravitational fragmentation into cores at 10 K. The filaments are systematically shorter in the cases with clustered feedback and are more closely distributed around the critical mass to length ratio for fragmentation.}
\label{fig:fil_props}
\end{center}
\end{figure}

Figure \ref{fig:fil_props} gives a more quantitative analysis of the filament properties for the different clouds. Note that the absolute value of the filament lengths will always depend on the method used. \disperse segments filaments every time there is a discontinuity (e.g. if a sink particle has eaten a large gap in the filament) or a sharp change in orientation (greater than 45 \textdegree), therefore the absolute values of the filament lengths are algorithm dependent. However, as the same methodology has been used for all our cloud complexes there is value in comparing the lengths. The top panel of Figure \ref{fig:fil_props} shows a histogram of the lengths of all the individual filaments shown by the coloured lines in Figures \ref{fig:fils_xy} and \ref{fig:fils_yz}. Continuous filaments of up to 40 pc are seen within the clouds in complexes A and B, however in the complexes with previous clustered feedback the filaments within the clouds are much shorter and rarely exceed 10 pc. As a consequence of these shorter lengths and lower cloud densities, the mass associated with each filament -- defined as gas within one filament width of the central filament spine as described in Section A -- is also lower in these cases (see middle panel in \ref{fig:fil_props}).

\begin{table}
	\centering
	\caption{Mean and median filament properties in each of the cloud complexes.}
		\begin{tabular}{c c c c c}
		\hline
	         \hline
	          Complex & \multicolumn{2}{c}{Mass [\msun]}  & \multicolumn{2}{c}{Length [pc]}  \\
	           & mean & med.\ & mean & med.\ \\
	   	 \hline
		 A & 598.5 & 324.3 & 9.4 & 5.8 \\
		 B & 539.9 & 523.1 & 9.7 & 5.3 \\
		 C & 34.6 & 25.6 & 4.4 & 3.2 \\
		 D & 35.8 & 20.1 & 3.9 & 2.9 \\
	         \hline	 
	         \hline        
		\end{tabular}
	\label{tab:filprops}
\end{table}

The bottom panel of Figure \ref{fig:fil_props} shows the mass to length ratios of the filaments in each complex. This is an important property as it determines how susceptible filaments are to fragmentation. \citet{Inutsuka97} showed that above a mass-to-length ratio of $16.7$ \msun pc$^{-1}$ an isothermal filament with a temperature of 10 K would fragment. Our gas is of course not isothermal, but we plot in grey this value on Figure \ref{fig:fil_props} for reference as a guide to the minimum mass needed for fragmentation. The majority of the filaments in both complexes A and B are likely highly unstable to fragmentation. This accounts for the large number of star-forming cores in Figures \ref{fig:fils_xy} and \ref{fig:fils_yz}. However, with clustered supernova feedback one naturally produces filament networks with a range of binding energies, meaning that only some of the filaments within the clouds are liable to fragmentation at a given time.

\begin{figure}
\begin{center}
\includegraphics[width=6.5cm]{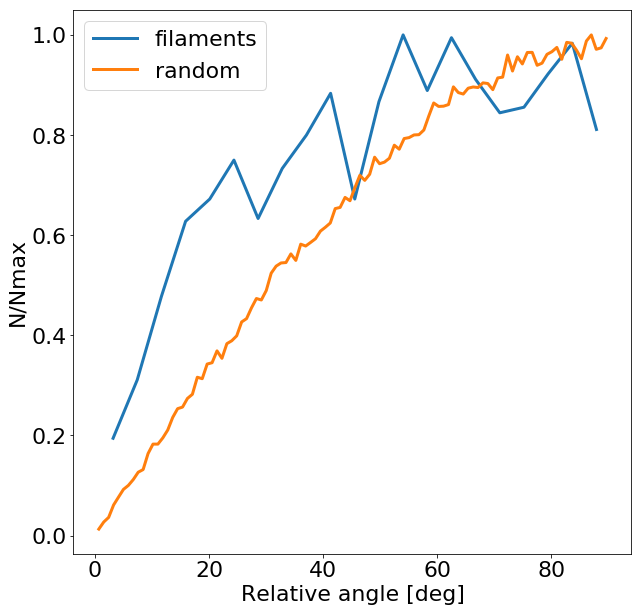}\\
\includegraphics[width=6.5cm]{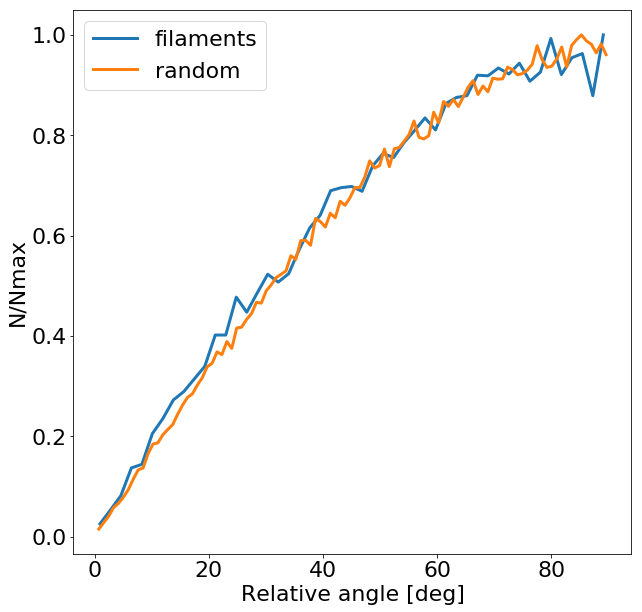}
\caption{The orientation of the filaments in a potential-dominated case (\textit{top}) and feedback-dominated case (\textit{bottom}).}
\label{fig:orientations}
\end{center}
\end{figure}

A further difference between the filaments in the two cases is in their orientations. As an example, Figure \ref{fig:orientations} shows the relative angle of the filaments with each other in the potential-dominated complex A and the clustered feedback dominated run D at a time 2 Myr after tracer particles were injected. The figure is normalised with respect to the maximum number of counts such that the distribution peaks at 1. The spine points along each filament skeleton are fitted with a single 3D vector, and then the angle with respect to every other filament is calculated. A relative angle of $0 \deg$ therefore then corresponds to parallel filaments and $90\deg$ to perpendicular filaments. The orange line shows a comparison where the angle between randomly orientated vectors is calculated. An excess in the number of parallel filaments is seen in complex A where the filaments follow the morphology of the spiral arms. In the clustered feedback dominated case, however, the filaments are consistent with being randomly orientated. This is a topic that we hope to return to in future papers when we focus on the effect of magnetic fields, which may also influence the alignment of filaments.

\subsection{Filament velocities}
\label{sec:cl_coherent}

\begin{figure}
\begin{center}
\includegraphics[width=6.5cm]{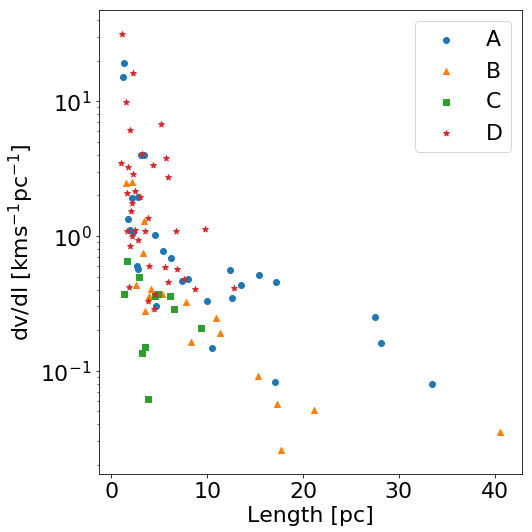}\\
\includegraphics[width=6.5cm]{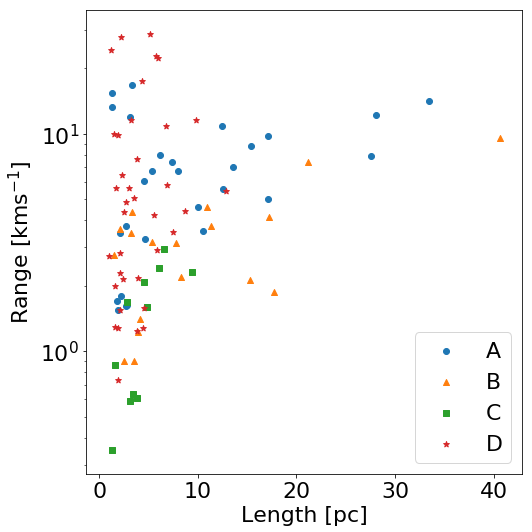}
\caption{The velocity gradients (\textit{top}) and velocity ranges (\textit{bottom}) of the filaments identified with \disperse in each of the cloud complexes.}
\label{fig:velocities}
\end{center}
\end{figure}

A major feature of observed filamentary clouds in the ISM is their velocity coherence. Figure \ref{fig:velocities} shows the gas velocity gradient along the filament spines. We calculate this by finding the mass-weighted mean velocity at each point along a filament spine, doing a linear regression to find the gradient of each velocity component, and then taking the magnitude of the three components. The gradient for each component is calculated over the entire filament length rather than taking an average of the gradient from point to point along the length. For nearly all filaments greater than a few pc the velocity gradients are under 1 \kms pc$^{-1}$. There is a clear tendency for longer filaments to have lower velocity gradients. This seems to be mainly a question of survival. Long filaments that are being rapidly destroyed by shear have a large velocity gradient. Figure \ref{fig:velocities} also shows the velocity ranges spanned by the filaments i.e. the difference between the magnitudes of the largest and the smallest velocity assigned to the filament. These show that in absolute terms the longest filaments do not span velocity ranges greater than 10 \kms. In our simulations low velocity gradients are a necessary consequence of the long filaments long-term survival.

\begin{table}
	\centering
	\caption{Mean filament velocity gradients of the resolved filaments in each of the cloud complexes in \kmspc. We show both the means for the full sample, but also for subsets of filaments of different lengths, $L$.}
		\begin{tabular}{l c c c c}
		\hline
	         \hline
	          Complex & All & $L<2$pc & 2pc$<L<$ 5pc & $L>5$pc \\
	   	 \hline
		 A &  2.88 & 14.62 & 2.04 & 0.47 \\
		 B &  0.79 & 3.97 & 1.12 & 0.20 \\
		 C &  0.86 & 1.34 & 0.96 & 0.26 \\
		 D &  2.48 & 3.90 & 2.23 & 1.22 \\
	         \hline	 
	         \hline        
		\end{tabular}
	\label{tab:filvels}
\end{table}

\begin{table}
	\centering
	\caption{ The parameters of a linear regression of filament length against the magnitude of the filament velocity gradients in each complex as shown in the first panel of Figure \ref{fig:velocities}. We also quote the standard error as found by the scipy linregress function and two sided p-value. In all cases there is a negative correlation between length and velocity gradient and this is most extreme for complexes A and D.}
		\begin{tabular}{l c c c c}
		\hline
	         \hline
	          Complex & slope & intercept & standard error & p-value \\
	   	 \hline
		 A &  -0.26 & 5.35 & 0.15 & 0.093 \\
		 B &  -0.05 & 1.31 & 0.02 & 0.030 \\
		 C &  -0.10 & 1.23 & 0.04 & 0.032 \\
		 D &  -0.37 & 3.91 & 0.10 & 0.0002 \\
	         \hline	 
	         \hline        
		\end{tabular}
	\label{tab:vel_regression}
\end{table}

Table \ref{tab:filvels} gives a more quantitative comparison of the filament velocity gradient between the different cloud complexes. Due to the strong change in gradient with filament length we give both the mean velocity gradient for the full complex, but also the gradient in filaments within three length ranges spanning this transition. The lowest velocity gradients occur in the inter-arm cloud complex B, which takes the form of a single continuous filament stretched out by differential rotation. The low feedback in this system has lead to an extremely coherent velocity structure. Surprisingly the two highest overall mean velocity gradients come from two very different complexes, A, which is the spiral arm with no clustered feedback, and D which is strongly feedback-dominated. However, a comparison of the gradients in filaments of different lengths shows there are clear differences between the two. 

In complex A some extremely high gradients (more than 10 \kmspc) are seen from short filaments that join the larger ones. These short filaments are quickly ripped apart as the cloud passes through the arm. However, the longer filaments have much lower velocity gradients allowing them to survive. A different picture emerges in complex D in which there are embedded supernova. Here the velocity gradient still decreases as the length increases, but much less sharply. In filaments of less than 2 pc long the gradient is less than a third of that in the spiral arm, but at lengths greater than 5 pc it is almost three times greater, and is still greater than 1 \kmspc unlike in any of the other clouds. Here the expanding supernovae bubble is acting on all the filaments to shear them apart (we will talk more about the cloud evolution in Section \ref{sec:sf}). Complex C represents an intermediate case, where the enhanced turbulence in the disc from supernova feedback has raised the general velocity dispersion, but there is no supernova feedback yet within the cloud.

This is further confirmed by performing a linear regression between all the filament lengths in each complex and the magnitude of their velocity gradients as shown in Table \ref{tab:vel_regression}. In all cases there is negative trend with the sharpest decrease seen in Complexes A and D. A linear trend towards decreasing velocity gradient with length is a good description of the data in complexes B, C and D, However it is a less good description of complex A, where the drop in velocity gradient between very small and intermediate filaments is extremely precipitous.

\subsection{Cloud evolution and star formation}
\label{sec:sf}

Finally we consider how the filament networks and star formation within the cloud complexes change as they evolve. As a reminder, our sink particles \textit{do not} represent individual stars, instead they represent sites where there is a collapsing clump/core of gas undergoing star formation. In future work we will seek to go closer to the sites of individual collapsing cores of gas that will form individual stellar systems to consider typical core masses and core formation efficiencies. But for now, it is still interesting to consider where sites of star formation are located. To allow the complexes to evolve for some time at the higher resolution, we start our analysis 1 Myr after the tracers were injected. We first focus on our two most extreme examples, cloud complex A, which resides in a spiral arm and does not have clustered feedback tied to sink particles, and cloud complex D which had pre-existing clustered feedback and is now also undergoing feedback from within.

\begin{figure*}
\begin{center}
\begin{tabular}{c c}

\begin{overpic}[scale=0.3]{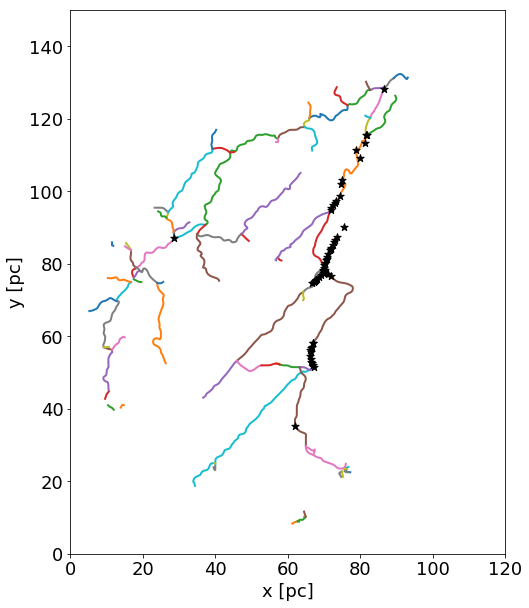}
\put (25,90) {\makebox(0,0){{\bf 1 Myr}}}
\end{overpic}
\begin{overpic}[scale=0.3]{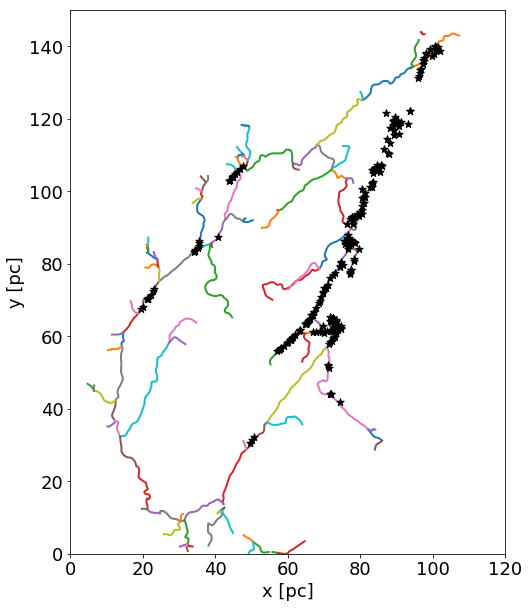}
\put (25,90) {\makebox(0,0){{\bf 2 Myr}}}
\end{overpic}
\begin{overpic}[scale=0.3]{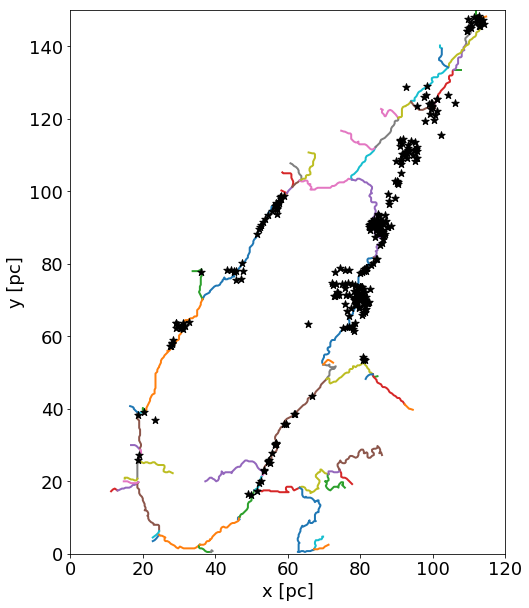}
\put (25,90) {\makebox(0,0){{\bf 3 Myr}}}
\end{overpic}

\end{tabular}
\caption{The location of the sites of star formation (shown with black stars) in cloud complex A with respect to the filament network at 1 Myr intervals after tracer refinement was turned on.}
\label{fig:starsA}
\end{center}
\end{figure*}

Figure \ref{fig:starsA} shows a schematic of where sinks are formed along the filaments in cloud complex A at 1 Myr intervals after the tracer refinement is turned on. As the filaments are extremely smooth and continuous due to the lack of clustered feedback they reach the threshold for fragmentation everywhere along their length almost simultaneously and turn into a line of collapsing clumps in a huge burst of star formation. These then interact dynamically to then become large clusters. These clusters seem to co-locate with the junctions of the filament network.

\begin{figure*}
\begin{center}
\begin{tabular}{c c c}

\begin{overpic}[scale=0.3]{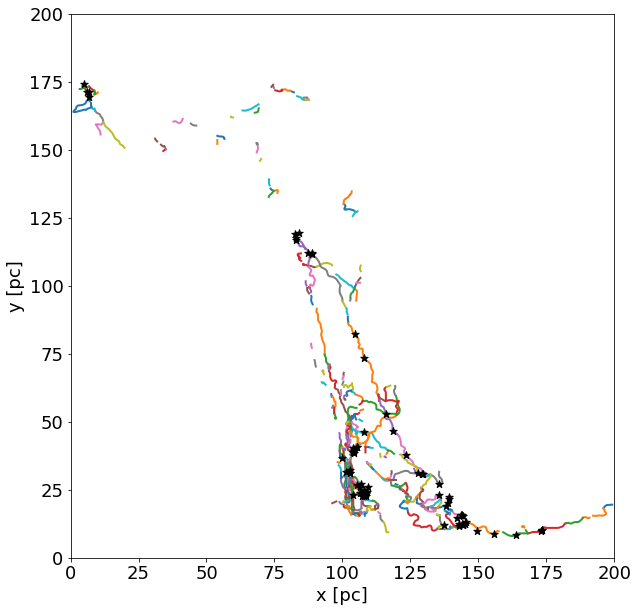}
\put (70,85) {\makebox(0,0){{\bf 1 Myr}}}
\end{overpic}
\begin{overpic}[scale=0.3]{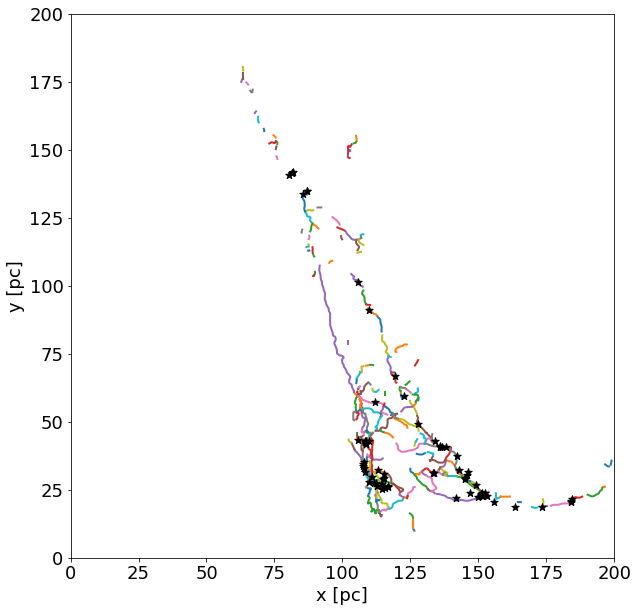}
\put (70,85) {\makebox(0,0){{\bf 1.5 Myr}}}
\end{overpic}\\
\begin{overpic}[scale=0.3]{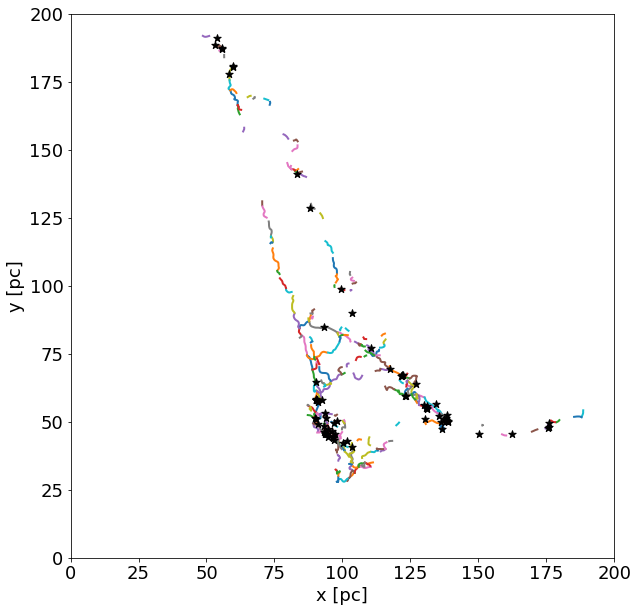}
\put (70,85) {\makebox(0,0){{\bf 2 Myr}}}
\end{overpic}
\begin{overpic}[scale=0.3]{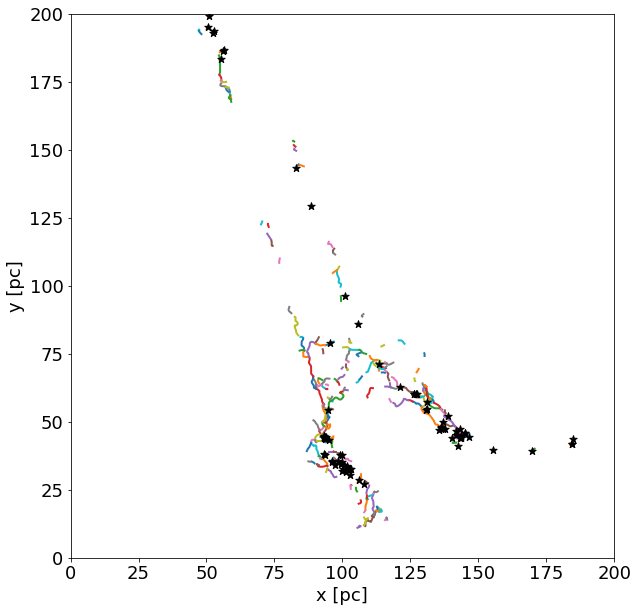}
\put (65,85) {\makebox(0,0){{\bf 2.5 Myr}}}
\end{overpic}
\end{tabular}

\caption{The location of the sites of star formation (shown with black stars) in cloud complex D with respect to the filament network at 0.5 Myr intervals after tracer refinement was turned on. We examine this network over a shorter period than Figure \ref{fig:starsA} as complex D is disrupted by feedback after 2.5 Myr.}
\label{fig:starsD}
\end{center}
\end{figure*}

Figure \ref{fig:starsD} shows the location of sinks in the strongly feedback-dominated case at 0.5 Myr intervals (as this is a more rapidly evolving region, we reduce the time separation between the images). In this case the filaments are shorter, less massive, and only a subset of them reach the critical $M/L$ ratio for fragmentation. This results in more distributed star formation. The first generation of stars can then produce feedback that starts to expel the remaining mass in the region and destroy the cloud. In the last time considered here (2.5 Myr) feedback from the bottom left corner has had a substantial impact upon the surrounding gas and after this point we find that the cloud and filament network has effectively dissolved.

\begin{figure*}
\begin{center}
\begin{tabular}{c c}
\begin{overpic}[scale=0.45]{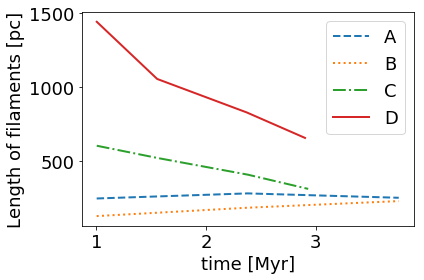}
\put (25,57) {\makebox(0,0){{\bf (a)}}}
\end{overpic}

\begin{overpic}[scale=0.45]{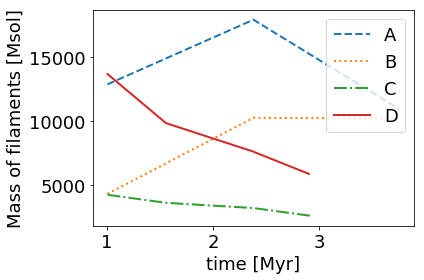}
\put (28,57) {\makebox(0,0){{\bf (b)}}}
\end{overpic}\\

\begin{overpic}[scale=0.45]{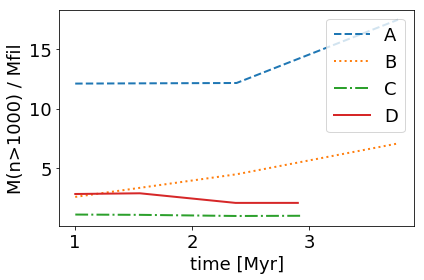}
\put (23,57) {\makebox(0,0){{\bf (c)}}}
\end{overpic}

\begin{overpic}[scale=0.45]{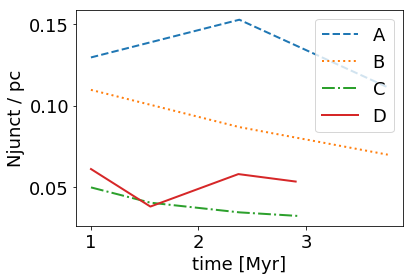}
\put (25,57) {\makebox(0,0){{\bf (d)}}}
\end{overpic}\\

\end{tabular}
\caption{The evolution of the cloud complexes and filamentary networks over time. \textit{(a)} The total length of resolved filaments in the simulated networks, \textit{(b)} the mass in resolved filaments, \textit{(c)} the dense gas ($n>1000$ \cmc) fraction in the cloud complex, and \textit{(d)} the ratio of junctions to the total length of filaments in pc.}
\label{fig:time_evo}
\end{center}
\end{figure*}

We can also investigate how the filament network within the cloud complexes evolves with time. Figure \ref{fig:time_evo} shows how various filament statistics evolve in each of the cloud complexes. The feedback-dominated complexes (C \& D) start out with far longer total lengths of resolved filaments within them (despite having less overall mass) but this falls over time as the feedback disrupts the filaments. Similarly the mass in resolved filaments also decreases. The potential-dominated complexes without the feedback start with a lower total length of resolved filaments, but this grows over time. The mass in resolved filaments in the potential-dominated regions also generally increases with time, with the exception of the final time period in complex A when the complex is slowly being stretched apart.

As both the mass and length of filaments are changing, we plot in the lower left panel of Figure \ref{fig:time_evo} the ratio of the mass in filaments vs the mass of dense gas within the cloud (arbitrarily defined as $n > 1000 \, {\rm cm^{-3}}$) to see how closely the two properties are connected. There is a higher ratio of dense gas per filament in the potential-dominated cases compared to the clustered feedback. Throughout the lifetimes of the cloud this ratio does not change much for each region.

Finally in the bottom right panel we investigate how the morphology of the cloud changes by calculating the number of junctions in the cloud filament networks. We define a junction as being a point in the network where three or more filaments join. This excludes locations where two sections that have been split due to a sink or a sharp change in direction but are otherwise continuous join together. As each network has a different total filament length we divide by the total filament length to get a fair comparison. A mixed picture arises. In complex A the number of junctions per unit length increases, but in B and C it decreases, whereas in complex D it remains constant or has no clear trend. In all cases the proportion of the filament length that is associated with a junction of filaments remains low, typically 0.05 - 0.1 junctions per parsec. This reinforces the premise that junctions in filament networks might be special places for star formation.

The filamentary nature of the star formation in these complexes also has implications for the morphology of the star clusters that arise from them, as the forming protostars will inherit the morphology and velocities of the gas that they form from. This is particularly significant given the low velocity gradients in our star-forming filaments. Recent \textit{Gaia} observations have indicated that there may be a population of star clusters that retain a filamentary geometry. For example, \citet{Kounkel19} used machine learning algorithms in \textit{Gaia} DR2 to identify new string-like groups of stars in the local group parallel to the galactic plane. Beccari, Boffin and Jerabkova (submitted) have used \textit{Gaia} DR2 to study the Vela OB2 region, finding a 260 pc wide 35 Myr old star cluster, which they interpret as a remnant of filamentary star formation.

\section{Discussion}
\label{sec:discussion}

\subsection{The role of large-scale forces and feedback in shaping cloud filament networks}
\label{sec:evaluation}

In this paper we have set-out to generate filament networks within molecular cloud complexes self-consistently taking into account the following large-scale forces that act outside the clouds from the galactic environment: 1) the galactic potential and spiral arms bringing gas together, 2) the wiggle instability, which causes the gaseous spiral arms to fragment, creating and amplifying the filaments \citep{Wada04,Sormani17a}, 3) differential rotation stretching clouds, and 4) random and clustered supernova feedback. This is in contrast to work in isolated cloud simulations, where filaments arise out of the turbulent field imposed in the initial condition. 

We analyse in detail four cloud complexes that are dominated by different forces. Complexes A and B are both dominated by the large-scale potential and rotation within the galaxy. They have random supernova feedback but not clustered feedback from sink particles, and self-gravity was not turned on before the tracer refinement. Complex A is located in a spiral arm and complex B in an inter-arm filamentary region. Complexes C and D were formed after a burst of clustered supernova feedback, which has disrupted material from the spiral arm. In complex C, star formation remains quite inefficient, but complex D has vigorous star formation that leads to internal supernova feedback that erodes its structure from within. 


The ISM properties, filament networks, and fragmentation within the cloud complexes formed in different galactic environments show substantial differences. In complexes A and B the filaments are systematically longer and more massive. Due to their smoothness they form sink particles almost simultaneously along the filament lengths, which then evolved into clusters that are typically associated with junctions in the filament networks. Complexes C and D which also include clustered feedback, have shorter and less massive filaments (and indeed the cloud complexes themselves are also less massive). The mass-to-length ratios of the filaments span a greater range of values, ranging from below to above the critical ratio, meaning that only a subset of the network is liable to fragmentation at any given time. The lower cloud densities and more sequential star formation make it easier for the cloud complex to be destroyed and the mass in filaments decreases after a few Myr.

In all cases the filaments have generally coherent velocities along their lengths. This is a necessary feature for their survival as large velocity gradients meant that the filament could not survive for an extended period. Large velocity gradients were only found in the shortest filaments. Generally, filaments longer than 2~pc had velocity gradients less than 1~\kmspc. An important exception was complex D, where supernova feedback occurring within the cloud increased the velocity gradients along long filaments and eventually destroyed them.

In reality our galaxy has clustered feedback and supernova bubbles and so we expect complexes C and D to be a more accurate depiction of real clouds. However, it is interesting to see how large an effect different formation mechanisms have on cloud filament networks. While analogues to Complexes A and B might be rarer, there are still occasions where they might be an important mode of star formation. For instance, complex A is formed from quiescent gas falling into a deep spiral arm potential. A good example of such a region might be the Nessie filament \citep{Jackson10,Goodman14,Zucker17} which is extremely long and coherent, lies in the galactic plane, and is closely associated with a spiral arm. Complex B has similarities with coherent low turbulence filaments such as the Musca filament \citep{Hacar16}, and in general with long filamentary clouds that are observed in inter-arm regions \citep[see e.g.][]{Ragan14}.

\subsection{Future work \& caveats}
\label{sec:future}

An important topic so far neglected is a comparison to observations. This is a deliberate oversight as such a comparison deserves a detailed treatment in its own right and must be done in the observational plane using radiative transfer to have any validity. \textit{Zucker et al. 2019 submitted} does a one-to-one comparison with observed non-self-gravitating filamentary clouds from \citet{Smith14a}, and we plan to extend this to the clouds studied here in the future. Similarly, work underway by \textit{Izquierdo et al. in prep} performs non-LTE line transfer of CO to investigate the turbulent properties of our cloud complexes. Several other works are in preparation or envisaged for the future, including work on magnetic fields, chemical evolution, star formation, and clustering.


A significant physical process not included in these simulations is magnetism. We know that both molecular clouds \citep[e.g.][]{Crutcher12} and galaxies \citep[e.g.][]{Beck13} have magnetic fields and this will affect how clouds are formed as well as the fragmentation within them. Its absence from these simulations is an important caveat and one that we are currently investigating using \arepo's magnetic field capabilities \citep{Pakmor11}. Another caveat that should be considered is our use of only supernova feedback, when we know that stellar winds, photoionisation and radiation pressure also play important roles in shaping the cold ISM. While these processes have been included before on galactic scales \citep{Hopkins12a} such simulations typically do not follow the cold molecular phase of the ISM in detail at these resolutions. For reasons of simplicity and computational efficiency, we have neglected these processes in our current simulations, but are working towards including these in future works. Our supernova feedback drives large bubbles in the ISM, this is both a consequence of there being an initial burst of feedback in the arms, but also due to the lack of early feedback from ionisation and winds. This means that some of the sinks can grow extremely massive and will have a large effect on the diffuse ISM if they become decoupled from the dense regions due to a dynamical event. The size of these bubbles is likely to decrease in future work if early feedback is included.

Finally, it would be interesting to investigate whether there is a significant difference in the widths of the filaments formed in the potential-dominated and feedback-dominated cases. However, although our resolution in the tracer refinement regions is extremely high for a galactic-scale simulation, it is still not yet high enough to be confident that the widths are fully converged. We have therefore refrained from any detailed width analysis in this work. In future we aim to go to yet higher resolutions and sink creation densities to study in depth the core mass function and star formation efficiencies in our different cloud complexes.

\section{Conclusions}
\label{sec:conclusions}

We have introduced a new suite of simulations, "The Cloud Factory", which self-consistently forms molecular cloud complexes at high enough resolution to resolve internal substructure all while including galactic-scale forces. We use a customised version of \arepo \citep{Springel10} that includes physics modules that allow for a detailed treatment of the cold molecular ISM. Important processes include time-dependent chemistry, H$_{2}$ self-shielding and dust attenuation of the interstellar radiation field, gas heating and cooling, sink particles representing regions of star formation, random supernova feedback, clustered supernova feedback tied to sink particles, and gas self-gravity. We use an idealised spiral arm potential and focus our analysis on the disc of the galaxy outside any bar. The goal of the calculations is not to fully model the evolution of a spiral galaxy like the Milky Way, but instead to simulate enough of the large-scale evolution to capture its influence on the cloud formation process. The simulations therefore act as a laboratory in which the impact of different forces on the cold molecular ISM can be studied.

We do not run the entire gas disc at the same resolution but instead increase our resolution to a target mass of 10 \msun \ within a 3 kpc box of the galaxy that co-rotates with the gas. From within this box we focus on interesting cloud complexes which we tag with tracer particles. In these tracer refinement regions we increase the resolution further to a target mass of 0.25 \msun \ (spatial scales of better than 0.1 pc at $n > 1000$~\cmc), but crucially the tracer refinement region continues to evolve within the galaxy simulation and is not "cut out", allowing us to properly capture the impact of long-range forces and the local galactic environment.

As an example of the power of the method, in this first paper we investigate the impact of galactic-scale forces on the filament networks formed within four very different cloud complexes: A) a spiral arm region with only random feedback dominated by the spiral potential; B) an inter-arm region with random feedback stretched by differential rotation; C) a cloud complex formed after a burst of clustered supernovae feedback in its vicinity has stirred the ISM; and D) a cloud complex formed after a burst of clustered feedback that then undergoes supernova feedback in its interior. Studying the networks of filamentary structures within such clouds is an important topic as filament fragmentation plays a key role in star formation and different filament characteristics will therefore lead to different fragmentation within the clouds. Such differences will be missed in isolated simulations of molecular clouds that neglect the large-scale formation mechanism and impose turbulence as part of their initial conditions.

We find that the filament properties in the cloud complexes are very different. Complexes A and B with only random supernova feedback have longer, more massive filaments that fragment simultaneously along their length. In complexes C and D where the effect of clustered supernovae is included the filaments are shorter and less massive, with a greater range of binding states meaning that only a subset of the network is liable to fragment at any given time. The filaments in complexes C and D are shorter-lived and more easily destroyed than the potential-dominated cases. There are also clear differences in filament alignment, with the filaments in A and particularly in B tending to be parallel to each other, whereas in C and D they are consistent with random orientations. The filaments tend to be velocity-coherent objects. Velocity gradients along the filament spines typically only exceed 1 \kmspc for short filaments of length less than 2 pc. The only exception to this is in complex D where supernova feedback within the cloud has increased the velocity gradients of the filaments and consequently disrupts them through shear. These filaments then go on to form stars which will inherit the properties of the gas from which they form, and may be the progenitors of recent filamentary clusters and groups of stars observed with \textit{Gaia} ( see e.g. \citealt{Kounkel19}.)

The clear differences between the cloud complexes ISM properties, the filament networks they form, and the fragmentation within them shows how galactic-scale forces have a real impact on star formation within molecular clouds. Such effects might lead to different star formation outcomes in spiral arm clouds such as Nessie \citep{Jackson10}, quiescent inter-arm filaments \citep{Ragan14}, and the more general case of clouds affected by previous supernova bubbles.

\section*{Acknowledgements}
We would like to thank the referee, Daniel Seifried, for constructive comments that improved the paper. RJS gratefully acknowledges support from an STFC Ernest Rutherford Fellowship (grant ST/N00485X/1), without which this work would not have been possible. MCS, SCOG, and RSK acknowledge financial support from the German Science Foundation (DFG) via the collaborative research centre (SFB 881) `The Milky Way System' (subprojects B1, B2, and B8) and from the Heidelberg cluster of excellence EXC 2181 `STRUCTURES: A unifying approach to emergent phenomena in the physical world, mathematics, and complex data' funded by the German Excellence Strategy. PCC and ADC acknowledge support from the Science and Technology Facilities Council (under grant ST/N00706/1). AFI acknowledges the studentship funded by the UK's Science and Technology Facilities Council (STFC) through the Radio Astronomy for Development in the Americas (RADA) project, grant number ST/R001944/1. C.Z. acknowledges support by NSF grant AST-1614941, ``Exploring the Galaxy: 3-Dimensional Structure and Stellar Streams.'' C.Z. is also supported by the NSF Graduate Research Fellowship Program (Grant No. 1650114) and the Harvard Data Science Initiative.

This work used the COSMA Data Centric system at Durham University, operated by the Institute for Computational Cosmology on behalf of the STFC DiRAC HPC Facility (www.dirac.ac.uk. This equipment was funded by a BIS National E-infrastructure capital grant ST/K00042X/1, DiRAC Operations grant ST/K003267/1 and Durham University. This research made use of SciPy \citep{Scipy}, NumPy \citep{Numpy}, and matplotlib, a Python library for publication quality graphics \citep{Matplotlib}.




\bibliographystyle{mnras}
\bibliography{Bibliography} 



\appendix

\section{Filament identification}
\label{sec:filid}
As discussed in the introduction, a key feature of the cold ISM is its filamentary nature. To identify filaments in the dense gas we use the \disperse (DIScrete PERsistent Structures Extractor) algorithm \citep{Sousbie11}. This constructs a Morse-Smale complex from an input density distribution and identifies the critical points where the density gradient is zero. Filamentary structures are found by connecting the points such that maxima are connected to saddle-points along Morse field lines. To avoid artifacts, we extract only the structures which have a persistence ratio with a probability of 5 sigma or more when compared to Poisson noise. We apply \disperse to a uniform grid of gas density that we generate from the above regions with a cell diameter of 0.5 pc (we will refer to this as the `finder grid' as it is only used to identify the structures, not to analyse them). We require that identified filaments must be above a minimum density threshold of 500 \cmc, thereby ensuring the gas is fully molecular and contains CO \citep{Smith14a}. We use this `skeleton' filament network to get a series of vectors describing the orientation of each filamentary structure.

The filament skeleton map obtained above does not contain any information about the properties of the filamentary structures, so to assign gas properties we have to calculate these from the \arepo simulation. We require a minimum of 10 \arepo cells per parsec along the length of the filament to call a filament resolved and include it in our analysis. This naturally means that we will exclude very diffuse filaments, and filaments that have been eaten away by sink particles when considering filament gas properties.

To estimate the filament mass we sum the mass of all \arepo cells within the estimated filament width of each vector that makes up a filament skeleton. Filament widths are known to vary along their length \citep{Suri19} and according to the definition used to define them \citep{Smith14b}. Here we adopt a simple prescription where we find the radial density profile along the length of each filament by calculating the shortest perpendicular distance of the \arepo gas cells to the filament spine and then plotting their density as a function of this distance. We set the width to be two times the radius from the filament centre to where the density falls below half the peak value (minus the background level) and do not allow the width to be shorter than half the pixel size of the finder grid used in \disperse (0.25 pc). The filament mass is assigned by tagging the cells within each filament width to find the gas belonging to it and then summing to get the total.

To assign velocities to the filaments we take the mass-weighted average at spine points along the filament skeleton of the gas perpendicular to the filament vector within this radius. This means that the gas properties of the filaments are calculated from the \arepo cells where the resolution is highest, not from the regular 3D finder grid used for identification with \disperse. Note that our filament identification and analysis is done purely in 3D, and not in the observational plane. To do this properly requires radiative transfer for the structures to be viewed inside the galaxy, and so we leave this for future work (see \textit{Zucker et al. submitted} for an example).


\bsp	
\label{lastpage}
\end{document}